\documentclass[twocolumn]{aastex631}
\usepackage{tablefootnote}
\usepackage{hyperref}
\usepackage{amsmath}

\begin{document}

\title{Improved constraints on the Faraday rotation towards eight fast radio bursts using dense grids of polarized radio galaxies}
\shorttitle{Precise magnetoionic environments of FRBs}

\author[0000-0002-8897-1973]{Ayush Pandhi}
\affiliation{David A. Dunlap Department of Astronomy and Astrophysics, University of Toronto, 50 St. George Street, Toronto, ON M5S 3H4, Canada}
\affiliation{Dunlap Institute for Astronomy and Astrophysics, University of Toronto, 50 St. George Street, Toronto, ON M5S 3H4, Canada}

\author[0000-0002-3382-9558]{B. M. Gaensler}
\affiliation{Department of Astronomy and Astrophysics, 1156 High Street, University of California Santa Cruz, Santa Cruz, CA 95064, USA}
\affiliation{David A. Dunlap Department of Astronomy and Astrophysics, University of Toronto, 50 St. George Street, Toronto, ON M5S 3H4, Canada}
\affiliation{Dunlap Institute for Astronomy and Astrophysics, University of Toronto, 50 St. George Street, Toronto, ON M5S 3H4, Canada}

\author[0000-0002-4795-697X]{Ziggy Pleunis}
\affiliation{Anton Pannekoek Institute for Astronomy, University of Amsterdam, Science Park 904, 1098 XH Amsterdam, The Netherlands}
\affiliation{ASTRON, Netherlands Institute for Radio Astronomy, Oude Hoogeveensedijk 4, 7991 PD Dwingeloo, The Netherlands}

\author[0000-0002-6952-9688]{Sebastian Hutschenreuter}
\affiliation{University of Vienna, Department of Astrophysics, Türkenschanzstrasse 17, 1180 Vienna, Austria}

\author[0000-0002-4119-9963]{Casey Law}
\affiliation{Cahill Center for Astronomy and Astrophysics, MC 249-17 California Institute of Technology, Pasadena, CA 91125, USA}
\affiliation{Owens Valley Radio Observatory, California Institute of Technology, Big Pine, CA 93513, USA}

\author[0000-0001-7348-6900]{Ryan Mckinven}
\affiliation{Department of Physics, McGill University, 3600 rue University, Montreal, QC H3A 2T8, Canada}
\affiliation{Trottier Space Institute, McGill University, 3550 rue University, Montreal, QC H3A 2A7, Canada}

\author[0000-0002-3968-3051]{Shane P. O'Sullivan}
\affiliation{Departamento de Física de la Tierra y Astrofísica \& IPARCOS-UCM, Universidad Complutense de Madrid, E-28040 Madrid, Spain}

\author[0000-0002-9822-8008]{Emily B. Petroff}
\affiliation{Perimeter Institute for Theoretical Physics, 31 Caroline St., Waterloo, ON, N2L 2Y5, Canada}

\author[0000-0001-7093-3875]{Tessa Vernstrom}
\affiliation{CSIRO, Space and Astronomy, PO Box 1130, Bentley, WA 6102, Australia}
\affiliation{ICRAR, The University of Western Australia, 35 Stirling Hw, Crawley, WA 6009, Australia}

\correspondingauthor{Ayush Pandhi}
\email{ayush.pandhi@astro.utoronto.ca}

\begin{abstract}
We present $2-4$~GHz observations of polarized radio galaxies towards eight fast radio bursts (FRBs), producing grids of Faraday rotation measure (RM) sources with sky densities of $9-28$ polarized sources per square degree. Using a Bayesian interpolation framework, we constrain Galactic RM fluctuations below $\sim 1~\mathrm{deg}^2$ angular scales around the FRB positions. Despite the positions of all eight FRBs far from the Galactic plane, we constrain previously unresolved small-scale Galactic RM structures around six of the eight FRBs. In two of these fields, we find potential changes in the sign of the Galactic RM that are not captured by previous, sparsely sampled RM grid observations. Our Galactic RM estimate towards the FRBs differs between a few$~\mathrm{rad}~\mathrm{m}^{-2}$ up to $\sim 40~\mathrm{rad}~\mathrm{m}^{-2}$ from the all-sky Galactic RM map of \cite{2022A&A...657A..43H}. Extrapolating our results to the known population of polarized FRB sources, we may be incorrectly interpreting the host galaxy RM for $\sim 30$\% of the FRB source population with current RM grid observations. Measuring small-scale Galactic RM variations is crucial for identifying FRBs in low density and weakly magnetized environments, which in turn could serve as potent probes of cosmic magnetism. This framework of reconstructing continuous Galactic RM structure from RM grid observations can be readily applied to FRBs that fall in the sky coverage of upcoming large-sky radio polarization surveys of radio galaxies, such as the Very Large Array Sky Survey (VLASS) and the Polarization Sky Survey of the Universe's Magnetism (POSSUM).
\end{abstract}


\section{Introduction} \label{sec:intro}
Fast radio bursts (FRBs) are millisecond-duration extragalactic radio transients that were discovered by \cite{2007Sci...318..777L}. To date, 802 FRB sources have been reported\footnote{Based on the \href{https://doi.org/10.7298/QFM2-YX83}{FRB Newsletter Volume 05, Issue 11} published in November 2024.} \cite[e.g., the first Canadian Hydrogen Intensity Mapping Experiment FRB (CHIME/FRB) catalog;][]{2021ApJS..257...59C} and just over $\sim 100$ of those sources have been associated with external galaxies \citep[e.g.,][]{2020ApJ...895L..37B, 2024ApJ...971L..51B, 2024Natur.635...61S, 2024arXiv240802083S}. FRBs have been observed to have a diverse set of local environments, including: in proximity to star forming regions \citep[FRB 20180916B;][]{2020Natur.577..190M}, in a globular cluster \citep[FRB 20200120E;][]{2022Natur.602..585K}, and associated with compact, persistent radio sources (PRSs) within dwarf galaxies \citep[FRB 20121102A and FRB 20190520B;][]{2017Natur.541...58C, 2022Natur.606..873N}.

In addition, the magnetoionic properties of the local environment enshrouding the FRB source are encoded in the Faraday rotation incurred by the linearly polarized emission of the FRB as it propagates to Earth. As linearly polarized emission propagates through magnetoionic media, it undergoes a wavelength squared ($\lambda^2$)-dependent rotation of its intrinsic polarization position angle $\psi_0$ to the observed angle:
\begin{equation}
\psi(\lambda) = \psi_0 + {\rm{RM}}\lambda^{2}\,. \label{eq:pa}
\end{equation} 
The rotation measure (RM) depends on the integrated number density of electrons and parallel component of the magnetic field along the line of sight (LOS):
\begin{equation}
{\rm{RM}} = 0.812 \int_{z}^{0} \frac{n_e(z) B_\parallel(z)}{(1 + z)^2}\frac{\mathrm{d}l(z)}{\mathrm{d}z}\mathrm{d}z~\mathrm{rad}~\mathrm{m}^{-2} \label{eq:RM} \,.
\end{equation}
In Equation \ref{eq:RM}, we integrate between the emitting source (at $z$) to the observer (at $z=0$) such that the RM is positive when the average LOS magnetic field is pointed towards the observer and negative when it is pointed away from the observer. The electron density ($n_e(z)$) is in units of cm$^{-3}$, the LOS magnetic field strength ($B_\parallel(z)$) is in units of $\mu$G, and $\mathrm{d}l(z)$ is the LoS line element at $z$ in units of pc. Propagation through the same medium also causes a $\lambda^2$ dispersion of the radio emission, which is parameterized as the dispersion measure (DM):
\begin{equation}
{\rm{DM}}= \int_{0}^{z} \frac{n_e(z)}{(1+z)} \frac{\mathrm{d}l(z)}{\mathrm{d}z}\mathrm{d}z~\mathrm{pc}~\mathrm{cm}^{-3} \label{eq:DM} \,.
\end{equation}

To precisely probe the local magnetoionic environment around the FRB, we need to account for, and subtract, contributions to the DM and RM from other, intervening, media along the LOS. We can express the observed DM and RM ($\mathrm{DM}_\mathrm{obs}$ and $\mathrm{RM}_\mathrm{obs}$) of an FRB in terms of the individual contributing components along the LOS:
\begin{equation}
\mathrm{DM}_{\mathrm{obs}} = \mathrm{DM}_{\mathrm{MW}} + \mathrm{DM}_{\mathrm{IGM}}(z) + \frac{\mathrm{DM}_{\mathrm{host}}}{(1 + z)}~\mathrm{pc}~\mathrm{cm}^{-3}\,; \label{eq:dm_comps}
\end{equation}
\begin{equation}
\mathrm{RM}_{\mathrm{obs}} = \mathrm{RM}_{\mathrm{ion}} + \mathrm{RM}_{\mathrm{MW}} + \mathrm{RM}_{\mathrm{IGM}}(z) + \frac{\mathrm{RM}_{\mathrm{host}}}{(1+z)^2}~\mathrm{rad}~\mathrm{m}^{-2}\,. \label{eq:rm_comps}
\end{equation}
In Equations \ref{eq:dm_comps} and \ref{eq:rm_comps}, the ``MW'' subscript denotes the Galactic DM or RM, which combines contributions from both the Milky Way (MW) disk and halo ($\mathrm{DM}_\mathrm{disk}$ and $\mathrm{DM}_\mathrm{halo}$ and $\mathrm{RM}_\mathrm{disk}$ and $\mathrm{RM}_\mathrm{halo}$), the ``IGM'' subscript is the collective DM or RM contribution from the intergalactic medium (IGM), and the ``host'' subscript encompasses DM or RM contributions from the FRB local environment and host galaxy. In Equation \ref{eq:dm_comps}, we have ignored the DM contribution from Earth's ionosphere, which is typically only $\sim 10^{-5}~\mathrm{pc}~\mathrm{cm}^{-3}$ \citep{2016ApJ...821...66L} but we include the ionospheric RM contribution ($\mathrm{RM}_{\mathrm{ion}}$) in Equation \ref{eq:rm_comps}, as $\mathrm{RM}_{\mathrm{ion}}$ is generally on the order of $\sim 0.1-1~\mathrm{rad}~\mathrm{m}^{-2}$ \citep{2019MNRAS.484.3646S}. To derive magnetoionic properties of FRB local environments and host galaxies, we are chiefly interested in obtaining $\mathrm{DM}_{\mathrm{host}}$ and $\mathrm{RM}_{\mathrm{host}}$. Thus, our goal is to precisely account for, and subtract, all other foreground components in Equations \ref{eq:dm_comps} and \ref{eq:rm_comps}.

In most cases, $\mathrm{RM}_{\mathrm{MW}}$ is a significant contributor in Equation \ref{eq:rm_comps}, and sometimes it can even be larger than the $\mathrm{RM}_{\mathrm{host}}$ contribution. The best estimate of $\mathrm{RM}_{\mathrm{MW}}$ over large areas of the sky is provided by \citet[][hereafter H22]{2022A&A...657A..43H}, who apply a Bayesian inference scheme on 55,190 RMs \citep[mostly from radio galaxies;][]{2023ApJS..267...28V}, to reconstruct an all-sky interpolated map of $\mathrm{RM}_{\mathrm{MW}}$. However, there are two major limitations with the H22 $\mathrm{RM}_{\mathrm{MW}}$ map: (i) the underlying RM data has an average density of only $\sim 1$ polarized source per square degree \citep[mostly using the data of][]{2009ApJ...702.1230T} and (ii) it has large statistical uncertainties (with typical uncertainties of $10-20~\mathrm{rad}~\mathrm{m}^{-2}$). Therefore, on average, RM variations on scales $\lesssim 1~\mathrm{deg}^2$ are unconstrained\footnote{The sky density of RM sources is very inhomogeneous, so the exact scale of RM variations that is resolved varies over the sky.}. 

When looking at the local magnetoionic environments around FRBs, the large statistical uncertainty in $\mathrm{RM}_{\mathrm{MW}}$ is propagated into our estimate for $\mathrm{RM}_{\mathrm{host}}$, which can make it difficult to, for example, identify FRBs originating from ``clean'' environments (i.e., with $\mathrm{RM}_{\mathrm{host}} \sim 0~\mathrm{rad}~\mathrm{m}^{-2}$). Identifying FRBs in these clean environments would provide evidence for FRB sources that lack dense circumburst material, which is a key prediction of many theories that invoke young neutron stars as FRB progenitors \citep[e.g.,][]{2016MNRAS.458L..19C, 2016MNRAS.462..941L, 2017ApJ...841...14M, 2019MNRAS.485.4091M}. Instead, these FRBs may indicate a population of sources that is older, where the circumburst environment has dissipated or the central engine has moved away from its point of origin over time. In addition, FRBs with $\mathrm{RM}_{\mathrm{host}} \sim 0~\mathrm{rad}~\mathrm{m}^{-2}$ can serve as direct probes of intergalactic magnetic fields and their evolution over cosmic time \citep[e.g., see][]{2016Sci...354.1249R, 2016ApJ...824..105A}.

Currently, the largest number of polarized FRB sources comes from the CHIME/FRB collaboration \citep[including 128 non-repeating and 41 repeating FRBs][]{2023ApJ...951...82M, 2024ApJ...968...50P, 2024arXiv241109045N} and these FRBs have typical localization uncertainties of $\lesssim 1~\mathrm{arcmin}$ \citep{Michilli2024}. Polarized FRBs detected using other telescopes often have even more precise sky positions than the CHIME/FRB sample, for example FRBs observed with the Deep Synoptic Array \citep[with $\lesssim 2~\mathrm{arcsec}$;][]{2024ApJ...967...29L} and the Australian Square Kilometer Array Pathfinder \citep[with $\lesssim 1~\mathrm{arcsec}$;][]{2019Sci...365..565B, 2020ApJ...901L..20B, 2020Natur.581..391M, 2024arXiv240802083S}. The current state-of-the-art H22 $\mathrm{RM}_{\mathrm{MW}}$ cannot accurately resolve RM variations on $\lesssim 1~\mathrm{deg}^2$ scales across the whole sky. For experiments requiring precise estimates of $\mathrm{RM}_{\mathrm{host}}$, such as studies of the magnetized IGM, we need higher density grids of polarized radio galaxies (i.e., RM grids) to probe RM variations on scales comparable to the FRB positional uncertainties.

In this paper, we aim to precisely constrain the Galactic RM towards eight FRBs, derive stronger constraints of their $\mathrm{RM}_{\mathrm{host}}$, and quantify the limitations imposed by the current H22 Galactic RM sky on our ability to identify FRBs from clean magnetoionic environments and use them as probes of cosmic magnetism. We present the results of two observing campaigns undertaken with the Karl G. Jansky Very Large Array (VLA) that detect RM grids of densities $9-28$ polarized sources per square degree, each in a region centered on one of eight known FRB sky positions. We apply a similar Bayesian interpolation scheme to the one used by H22, but designed to be applied on discretized patches of the sky \citep{affan}, with which we reconstruct high resolution maps of $\mathrm{RM}_{\mathrm{MW}}$ toward each FRB. We compare our results to those by H22 and discuss the small-scale RM fluctuations that we are now able to probe with higher density RM grids. Using these maps, we obtain an updated $\mathrm{RM}_{\mathrm{MW}}$ towards each of the eight FRBs and more precisely constrain their local magnetoionic environments. We contextualize these results in light of upcoming large-sky polarization surveys and a rapidly growing FRB catalog to highlight key science that will become possible within the next few years.

The remainder of this paper is structured as follows. In Section \ref{sec:data} we summarize the two VLA observing campaigns, the data reduction, and the compilation of RM grids. In Section \ref{sec:foregrounds}, we break down the different DM and RM LOS components and our methodology of estimating each one. Results presenting the new $\mathrm{RM}_{\mathrm{MW}}$ reconstructions and their impact on the FRB LOS RM decomposition are provided in Section \ref{sec:results}. We discuss our results in Section \ref{sec:discussion} and identify key science questions that can be tackled with our methodology in the coming years.

\section{Data processing} \label{sec:data}
The VLA is a radio telescope based in the Plains of San Agustin, New Mexico, and operated  by the National Radio Astronomy Observatory (NRAO). The telescope is an array of 27 antennas; each antenna carries 10 receivers and is a 25-meter parabolic dish installed on an altitude-azimuth mount. The antennas are arranged into three elongated arms that each contain 9 antennas, allowing for a high degree of flexibility and control over baselines. The VLA has 4 standard configuration sizes: A, B, C, and D, which have longest baselines of 36.4, 11.4, 3.4, and 1.0~km, respectively.

\subsection{2018 VLA Observations} \label{sec:2018_vla}
The first of two observing campaigns was conducted between February 22-27, 2018 (PI: Dr.~Jamie Farnes) around FRB 20110523A \citep[for more details about this FRB source, see][]{2015Natur.528..523M}. These data were taken in the B-configuration at $2.0 -4.0 $~GHz with $2$~MHz channel spacing, $8$-bit samplers, an integration time of $3$~seconds and contain full polarization information. The observations have a Stokes I root-mean-square (rms) sensitivity of $\sim 30.7~\mu\mathrm{Jy}/\mathrm{beam}$ and the synthesized beamwidth is $2.1''$. The observation targeted $\sim 90$ known radio sources from the National Radio Astronomy Observatory VLA Sky Survey \citep[NVSS;][]{1998AJ....115.1693C} within a $1~\mathrm{deg}$ radius of FRB 20110523A. Each radio source was observed for $\sim 2.5~\mathrm{minutes}$, with the overall desired RM grid density of $\sim 25~\mathrm{sources}~\mathrm{deg}^{-2}$. The flux density scale, bandpass, delay, and polarization angle are calibrated using the bright radio source 3C286. The instrumental polarization leakage was calibrated with a known low polarization source, J1407+2827, and the complex gain calibrator was J2136+0041.

This general observing strategy was adopted and modified for follow-up RM grid observations around seven other FRBs in 2023, described in the next section.

\subsection{2023 VLA Observations} \label{sec:2023_vla}
The second observing campaign (PI: Ayush Pandhi) was undertaken between February 18 and May 23, 2023 while the telescope was in its B-configuration. These radio continuum observations were also taken at $2.0 - 4.0$~GHz with $2$~MHz channel spacing, $8$-bit samplers, an integration time of $3$~seconds, an on source time of $\sim 4~\mathrm{minutes}$ per target, and we retain full polarization information. The Stokes I rms sensitivity is consistently $18.75~\mu\mathrm{Jy/beam}$ across all observed fields and the synthesized beamwidth is $2.1''$. For these observations we target a total of $164$ radio sources in seven fields around FRBs 20160102A, 20181030A, 20190117A, 20190213B, 20190608B, 20191108A, and 20200120E.

The average target RM grid density around each of the seven FRBs was $\sim 25$~polarized sources per square degree. Thus, the average separation between RM grid sources approximately corresponds to the width of the error circle of the poorest localized FRB in our sample, FRB 20160102A. The RM grids are comprised of radio sources that were identified in the VLA Sky Survey \citep[VLASS;][]{2020PASP..132c5001L} quick look catalog \citep[][]{2021ApJS..255...30G}. VLASS is observed in the same frequency range ($2.0-4.0$~GHz) and has the same angular resolution ($2.1''$) as our VLA observations, though a complete polarization catalog from the survey has not yet been released. Selecting our RM grid sources from this catalog ensures that we are confident in the flux density of each radio source and know a-priori whether sources will be resolved or unresolved. From this catalog, we targeted all $\geq 3~\mathrm{mJy}$ radio sources in a $0.5$~degree radius around each FRB position. Since most of the radio sources that we target are within a few mJy in flux of our faintest source, we observe all targets for $\sim 4$~minutes. Importantly, we diverge from the observing strategy employed in 2018 by selecting fewer but brighter radio sources per target and increasing the time on source. This was done with the expectation that a higher fraction of targeted sources would yield sufficient linearly polarized signal to measure their RMs. We only conduct on-axis observations of these radio galaxies in order to avoid off-axis instrumental polarization effects, which can introduce $\sim 10$\% RM uncertainties \citep{2019MNRAS.487.3454M}.

For each field, we observed the quasar 3C138 as our flux density scale, bandpass, delay, and polarization angle calibrator. The instrumental leakage terms were calibrated by observing 3C84 -- a bright, low polarization ($< 1$\%) source. A complex gain calibrator near each target field was observed every $\sim 12$~minutes; for this calibration we used: J2248-3235 for FRB 20160102A, J1048+7143 for FRBs 20181030A and 20200120E, J2152+1734 for FRB 20190117A, J1800+7828 for FRB 20190213B, J2246-1206 for FRB 20190608B, and J1037+3309 for FRB 20191108A.

\subsubsection{FRB sample} \label{subsec:frb_data}
Multiple criteria were considered during the selection of FRB fields for the 2023 VLA observing campaign. First, we want to select FRBs for which the $\mathrm{RM}_\mathrm{MW}$ contribution dominates, or is comparable to, the $\mathrm{RM}_{\mathrm{host}}$, as these FRBs most benefit from our analysis. We enforce this criteria by requiring that the FRB must already have a measured $\mathrm{RM}_\mathrm{obs}$ and it must be $|\mathrm{RM}_\mathrm{obs}| < 1000~\mathrm{rad}~\mathrm{m}^{-2}$ since $\mathrm{RM}_\mathrm{MW} \ll 1000~\mathrm{rad}~\mathrm{m}^{-2}$ except along specific sightlines through the Galactic plane. We avoid FRBs near the Galactic plane by requiring that they must have a Galactic latitude $|b| > 25~\mathrm{deg}$ to minimize the likelihood of sharp $\mathrm{RM}_\mathrm{MW}$ gradients that may be difficult to model even with high density RM grids. Priority is given to FRBs with positional uncertainties smaller than the sky area sampled by the currently available all-sky RM grid of density $\sim 1$ polarized source per square degree. Ideally we want our RM grids to probe a comparable angular scale to the FRB positional uncertainty. Furthermore, FRBs with associated host galaxies and independently measured redshifts are also prioritized. This information allows us to measure the $\mathrm{RM}_{\mathrm{host}}$ contribution in the rest frame of the FRB and estimate the average LOS magnetic field strength in the FRB host galaxy. Finally, we deliberately choose to include a mix of repeating and non-repeating FRBs as target fields to ensure diversity in our sample and to enable potential analysis comparing the two sub-populations.  Note that the selection process for these FRB fields predates some of the recent FRB polarization catalogs \citep[e.g.,][]{2024ApJ...964..131S, 2024ApJ...968...50P}, and hence the candidate pool of FRBs that satisfied the aforementioned criteria at the time these observations were proposed was very limited. Based on these criteria, we selected seven FRBs around which we aimed to derive high density RM grids using targeted VLA observations. 

\begin{figure*}
    \centering
    \includegraphics[width=0.98\textwidth]{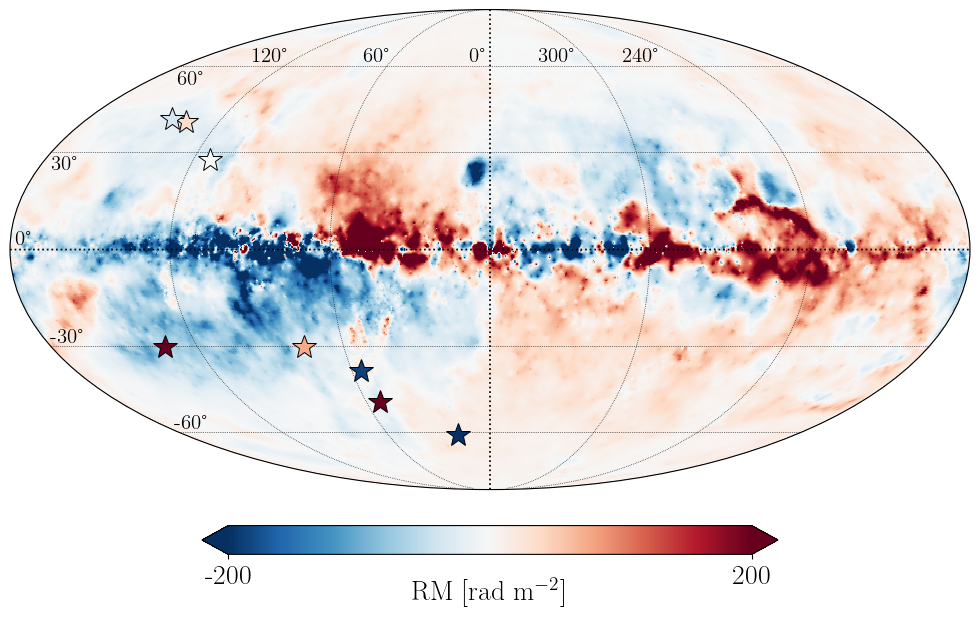}
    \caption{The eight targeted FRBs (stars) overlaid on the H22 reconstructed Galactic RM sky. Both the FRB markers and background map are colored based on their RM. For repeating FRBs, the mean RM across all bursts is used. The colorbar is saturated at $\pm 200~\mathrm{rad}~\mathrm{m}^{-2}$.} 
    \label{fig:frb_locs}
\end{figure*}

The seven FRBs in the 2023 VLA observations, as well as FRB 20110523A from the 2018 observations, have properties listed in Table \ref{tb:frbs}. In Figure \ref{fig:frb_locs} we show their sky positions, colored by their $\mathrm{RM}_\mathrm{obs}$, overlaid on the H22 $\mathrm{RM}_\mathrm{MW}$ sky.

\setlength{\tabcolsep}{1.5pt}
\setlength{\LTcapwidth}{1.0\textwidth}
\begin{table*}
\begin{center}
\caption{Summary of observed properties of the eight FRB sources (with repeaters in bold face) discussed in this work.} \label{tb:frbs}
\begin{tabular}{ccccccc}
\hline
\hline
TNS Name & Modified & J2000 Right  & J2000  & $\mathrm{DM}_\mathrm{obs}$ & $\mathrm{RM}_\mathrm{obs}$ & $z_\mathrm{host}$\\
 & Julian Date$^\mathrm{a}$ & Ascension & Declination & (pc~cm$^{-3}$) & (rad~m$^{-2}$) & \\
\hline
FRB 20110523A$^{1}$ & $55704.6$ & $21\mathrm{h}45\mathrm{m}12\mathrm{s}^\mathrm{b}$ & $-00\mathrm{^\circ}09\mathrm{m}37\mathrm{s}^\mathrm{b}$ & $623.30(6)$ &  $-186.1(1.4)$ & $-$\\
FRB 20160102A$^{2,3}$ & $57389.4$ & $22\mathrm{h}38\mathrm{m}49\mathrm{s}$ ($7.5'$) & $30\mathrm{^\circ}10\mathrm{m}50\mathrm{s}$ ($7.5'$) & $2596.1(3)$ &  $-221(6.4)$ & $-$\\
\textbf{FRB 20181030A}$^{4,5}$ & $58870.4$ & $10\mathrm{h}34\mathrm{m}22\mathrm{s}$ ($30.6''$) & $73\mathrm{^\circ}45\mathrm{m}14\mathrm{s}$ ($47''$) & $103.500(13)$ &  $36.39(29)$ & $0.00385$\\
 & $58870.4$ &  &  & $103.504(11)$ &  $37.67(54)$ & \\
 & $58870.9$ &  &  & $103.493(83)$ &  $37.08(51)$ & \\
\textbf{FRB 20190117A}$^{5,6}$ & $58500.9$ &  $22\mathrm{h}06\mathrm{m}38\mathrm{s}$ ($13''$) & $17\mathrm{^\circ}22\mathrm{m}06\mathrm{s}$ ($13''$) & $393.062(74)$ &  $74.30(68)$ & $-$\\
 & $59532.1$ &  &  & $395.804(36)$ &  $76.31(40)$ & \\
\textbf{FRB 20190213B}$^{5,6}$ & $58834.4$ & $18\mathrm{h}25\mathrm{m}02\mathrm{s}$ ($20''$) & $81\mathrm{^\circ}24\mathrm{m}05\mathrm{s}$ ($26''$) & $301.64(12)$ &  $-3.63(41)$ & $-$\\
 & $59055.2$ &  &  & $301.38(22)$ &  $1.04(42)$ & \\
 & $59261.7$ &  &  & $301.400(48)$ &  $1.06(52)$ & \\
 & $59314.0$ &  &  & $301.720(64)$ &  $0.97(52)$ & \\
 & $59431.3$ &  &  & $301.430(19)$ &  $-0.28(42)$ & \\
FRB 20190608B$^{7,8}$ & $58643.0$ & $22\mathrm{h}16\mathrm{m}05\mathrm{s}$ ($0.2''$) & $-07\mathrm{^\circ}53\mathrm{m}54\mathrm{s}$ ($0.2''$) & $340.05(6)$ &  $353(2)$ & $0.11778$\\
FRB 20191108A$^{9}$ & $58795.8$ & $01\mathrm{h}33\mathrm{m}47\mathrm{s}$ ($3.5'$)$^\mathrm{c}$ & $31\mathrm{^\circ}51\mathrm{m}30\mathrm{s}$ ($3.5'$) & $588.1(1)$ &  $474(3)$ & $-$\\
\textbf{FRB 20200120E}$^{10,11,12}$ & $58868.4$ & $09\mathrm{h}57\mathrm{m}54.69935\mathrm{s}$ ($1.2~\mathrm{mas}$) & $68\mathrm{^\circ}49\mathrm{m}00.8529\mathrm{s}$ ($1.3~\mathrm{mas}$) & $87.782(3)$ & $-29.8(5)$ & $^\mathrm{d}$\\
 & $59182.6$ &  &  & $87.812(4)$ &  $26.8(3)$ & \\
 & $59265.9$ &  &  & $87.7527(3)$ &  $-21.9(13.1)$ & \\
 & $59265.9$ &  &  & $87.7527(3)$ &  $-57.2(5.1)$ & \\
 & $59280.7$ &  &  & $87.7527(3)$ &  $-37.1(4.2)$ & \\
 & $59280.8$ &  &  & $87.7527(3)$ &  $-36.4(8.0)$ & \\
\hline
\hline
\multicolumn{7}{l}{$^{1}$ \cite{2015Natur.528..523M}; $^{2}$ \cite{2018MNRAS.475.1427B}; $^{3}$ \cite{2018MNRAS.478.2046C}; $^{4}$ \cite{2021ApJ...919L..24B}; $^{5}$ \cite{2023ApJ...951...82M};} \\
\multicolumn{7}{l}{$^{6}$ \cite{2023ApJ...950..134M}; $^{7}$ \cite{2020MNRAS.497.3335D}; $^{8}$ \cite{2021ApJ...922..173C}; $^{9}$ \cite{2020MNRAS.499.4716C}; $^{10}$ \cite{2021ApJ...910L..18B};} \\
\multicolumn{7}{l}{$^{11}$ \cite{2022Natur.602..585K}; $^{12}$ \cite{2022NatAs...6..393N}} \\
\multicolumn{7}{l}{$^{\mathrm{a}}$ The reported time of arrival of each burst at the respective observing telescope.} \\
\multicolumn{7}{l}{$^{\mathrm{b}}$ No positional uncertainties are reported.} \\
\multicolumn{7}{l}{$^{\mathrm{c}}$ The localization uncertainty was presented for the semi-major and semi-minor axes, which do not correspond to the} \\
\multicolumn{7}{l}{Right ascension and Declination uncertainties directly. To be conservative, we have presented the uncertainty on the semi-} \\
\multicolumn{7}{l}{major axis as the uncertainty on both the Right ascension and Declination.} \\
\multicolumn{7}{l}{$^\mathrm{d}$ This FRB source is associated with the M81 galaxy, which is at a distance of $3.6~\mathrm{Mpc}$.} \\
\end{tabular}
\end{center}
\end{table*}

\subsection{VLA data reduction} \label{subsubsec:data_reduc}
All eight fields were processed through the standard VLA calibration pipeline,\footnote{Details about the pipeline are \hyperlink{https://science.nrao.edu/facilities/vla/data-processing/pipeline}{provided by the NRAO}.} which performs basic flagging and calibration in the Common Astronomy Software Applications (CASA) program \citep{2022PASP..134k4501C}. For each field, we performed additional manual processing in CASA for each $128$~MHz spectral window separately. First, we applied Hanning smoothing across the band to mitigate any radio frequency interference (RFI) that was not initially flagged, removed antenna-shadowed data, and removed RFI outliers in the time-frequency space using the {\tt TFCrop} algorithm (using the default $4\sigma$ and $3\sigma$ thresholds for the time and frequency axes, respectively). The model visibility amplitude and phases of our calibrators were set using existing models of standard calibrators or by the model generated by the VLA calibration pipeline. We performed an initial phase and bandpass calibration on a subset of the central channels to average over temporal variation in the phases. Using the bandpass calibrator, we then solved for the antenna-based delays relative to the reference antenna and derive a bandpass solution. For the polarization calibration, we first solved for the residual delay between the two cross-hand polarization outputs using 3C138, then calibrated the instrumental polarization using the unpolarized source 3C84, and solved for the correct polarization position angle using the reference model of 3C138. Finally, we derived the complex antenna gain corrections across our observation, setting the zero phase to that of the reference antenna, using the respective gain calibrator for each field. The full set of calibration solutions was applied to all observed sources and we used the {\tt TFCrop} and {\tt RFFlag} algorithms for one last stage of RFI cleaning.

To clean and image our data, we employed the {\tt tclean} functionality of CASA using an adapted version of the {\tt CLEAN} deconvolver \citep{1974A&AS...15..417H}. We defined the reference frequency to be $3.0$~GHz, used two Taylor coefficients in the spectral model, applied a \cite{1995PhDT.......238B}  weighting scheme with the ``robust'' parameter set to $0.5$, and set the maximum number of total iterations to 20,000 with a maximum number of minor-cycle iterations to be 1,000. For every observed radio source, we produced both a multi-frequency synthesis image (i.e., combining the signal over the observing frequency range to produce a single image) and Stokes $I$ (total intensity), $Q$, $U$ (orthogonal linear polarizations), and $V$ (circular polarization) data cubes. 

\subsubsection{RM-synthesis} \label{subsubsec:rm_synth}
To estimate the RM of each radio source, we perform RM-synthesis \citep[as implemented in the {\tt RM-tools} package;][]{2020ascl.soft05003P} along the LOS corresponding to its brightest pixel. Note that the {\tt RM-tools} implementation also accounts for the spectral index of the radio source by modeling the Stokes $I$ spectrum and dividing it from the Stokes $Q$ and $U$ spectra. Here, we briefly summarize the RM-synthesis algorithm but for a more detailed description we refer the reader to the work by \cite{1966MNRAS.133...67B} and \cite{2005A&A...441.1217B}. RM-synthesis relates the complex linearly polarized intensity, $P(\lambda^2) = Q + iU$, to the Faraday dispersion function (FDF), $F(\phi)$, through a Fourier like transformation:
\begin{equation}
F(\phi) = \int_{-\infty}^{\infty}P(\lambda^2)e^{-2i\phi\lambda^2}\rm{d}\lambda^2 \,. \label{eq:fdf}
\end{equation}
Here, $\phi$ is the Faraday depth and is a generalized form of the RM that accounts for multiple sources of Faraday rotation in the polarized emission along the LOS. In practice, RM-synthesis computes $F(\phi)$ for a range of trial values of $\phi$ to measure the total linearly polarized intensity across the observed spectrum at different Faraday depths. The limited bandwidth and sampling of the data means that the FDF is convolved with the rotation measure spread function \citep[RMSF;][]{2005A&A...441.1217B}. The full width at half maximum (FWHM) of the RMSF, $\delta \phi$, effectively determines the resolution of the FDF in Faraday depth space as:
\begin{equation}
\delta\phi = \frac{3.79}{(\lambda_\mathrm{max}^2 - \lambda_\mathrm{min}^2)}\,, \label{eq:rmsf}
\end{equation}
for the maximum ($\lambda_\mathrm{max}$) and minimum ($\lambda_\mathrm{min}$) observed wavelength, respectively. For our observations between $2-4$~GHz, we achieve $\delta\phi = 224.9~\mathrm{rad}~\mathrm{m}^{-2}$.

Faraday rotation contributed by an intervening screen appears as a delta function at $\phi=\mathrm{RM}_\mathrm{obs}$ convolved with the RMSF. Previous works have found that polarization properties derived with RM-synthesis may become unreliable for a polarized signal-to-noise ratio of $(\mathrm{S/N})_P \leq 7$ \citep{2005A&A...441.1217B, 2012ApJ...750..139M}. Following the conservative approach used in a recent RM data release by the Polarization Sky Survey of the Universe's Magnetism (POSSUM) collaboration \citep{2024AJ....167..226V}, we consider a RM detection for a given radio source to be robust if it satisfies $(\mathrm{S/N})_P = 8$. In this case, the $\phi$ value corresponding to the center of the peak in the FDF is the estimated RM. The measurement uncertainty on the $\mathrm{RM}_\mathrm{obs}$, $\delta(\mathrm{RM}_\mathrm{obs})$, is:
\begin{equation}
\delta(\mathrm{RM}_\mathrm{obs}) = \frac{\delta\phi}{2(\mathrm{S/N})_P}\,. \label{eq:rm_err}
\end{equation}

\subsubsection{Constructing the RM grids} \label{subsubsec:rm_grids}
After applying RM-synthesis to all observed radio sources, we can begin constructing the respective RM grids for each field. Keeping in mind that the purpose of these RM grids is to inform our view of $\mathrm{RM}_\mathrm{MW}$, we want to ensure that each radio source provides a clean probe of our Galaxy and, thus, we want to minimize contamination from other contributions along the LOS. We classify radio sources into three categories based on their FDF: (i) unpolarized, (ii) ``Faraday simple'', or (iii) ``Faraday complex''. In the foremost case, there is no $(\mathrm{S/N})_P > 8$ peak in the FDF, so we are unable to sufficiently characterize its linear polarization properties and no RM estimate is derived. Unpolarized radio sources are removed from our sample and do not contribute to any RM grids. Faraday simple radio sources are those that have a singular peak in the FDF with $(\mathrm{S/N})_P > 8$ at $\phi = \mathrm{RM}$ and their RM uncertainty is computed following Equation \ref{eq:rm_err}. All Faraday simple sources are included in our RM grids. We note here that the determination of whether a linearly polarized radio source is Faraday simple or complex depends on whether we can resolve the different FDF components given our RMSF; hence, this determination depends on the frequency range of our observations. For example, a source that appears Faraday simple at high frequencies (due to a large RMSF that cannot resolve multiple, closely spaced FDF components) may actually be Faraday complex at low frequencies where the RMSF is smaller.

In some cases, the FDF of radio sources will have a $(\mathrm{S/N})_P > 8$ detection but deviate from the single peak expected from a Faraday simple source. These sources are labelled ``Faraday complex'' and require careful handling if they are to be used in an RM grid to probe a foreground screen. Faraday complexity can manifest as multiple FDF peaks or broadening of a FDF peak due to a variety of physical conditions, some examples include: multiple distinct RM contributing media along the LOS, unresolved RM components within the synthesized beam, an extended medium along the LOS that is both Faraday rotating and synchrotron emitting, or turbulence in a foreground magnetoionic screen on scales smaller than the projected size of the polarized source \citep[for a more detailed description, see][]{2024AJ....167..226V}. When dealing with multiple peaks in the FDF, it is difficult to ascribe a single RM value to the Faraday rotating screen that one is attempting to constrain. Meanwhile, for broadened FDF peaks, Equation \ref{eq:rm_err} may not be an accurate description of the RM measurement uncertainty.

To quantify the extent of Faraday complexity present in a radio source's FDF, we employ the $M_2$ metric developed by \cite{brown2011_m2}. First, we run the {\tt RM-CLEAN} algorithm \citep{2009A&A...503..409H} to deconvolve the observed FDF with the RMSF and obtain a list of discrete $\phi$ values and their $|F(\phi)|$ amplitudes that exceed a signal-to-noise threshold of $8\sigma$. We purposely choose $8\sigma$ to match our $(\mathrm{S/N})_P > 8$ threshold following the setup used by \cite{2024AJ....167..226V} to avoid over-cleaning the data. Using the {\tt RM-CLEAN} products as inputs, we can then compute $M_2$ as the square root of the second moment of $\phi$:
\begin{equation}
M_2 = \left[ \frac{\sum_{i=1}^{N} (\phi_i - \overline{\phi})^2 |F(\phi_i)|}{\sum_{i=1}^{N} |F(\phi_i)|} \right]^{1/2}\,, \label{eq:m2}
\end{equation}
where the $i$ subscript denotes each $> 8\sigma$ {\tt RM-CLEAN} component for a total of $N$ components and $\overline{\phi}$ is the first moment:
\begin{equation}
\overline{\phi} = \frac{\sum_{i=1}^{N} \phi_i |F(\phi_i)|}{\sum_{i=1}^{N} |F(\phi_i)|}\,. \label{eq:phi_i}
\end{equation}
To account for our effective resolution in $F(\phi)$ space, we normalize $M_2$ by $\delta\phi$,
\begin{equation}
m_2 = \frac{M_2}{\delta\phi}\,. \label{eq:m2_norm}
\end{equation}
\citep{2024AJ....167..226V} and use $m_2$ as our final complexity metric. 

For a Faraday simple source, with only one $> 8\sigma$ {\tt RM-CLEAN} component, we would measure $m_2 = M_2 = 0$. As one adds more {\tt RM-CLEAN} components, with increasing separation between them, and as their $|F(\phi)|$ amplitude becomes comparable to the primary FDF peak, $m_2$ increases. There are no strict rules regarding setting complexity thresholds for $m_2$, so we determine two thresholds upon visual inspection of our data. First, we remove any radio sources with $m_2 > 0.25$ from our sample, as these correspond to sources with multiple resolved peaks in their FDF such that we cannot unambiguously determine the Galactic RM along the LOS. We decide to include linearly polarized radio sources with $0.1 < m_2 < 0.25$, and with at least one secondary {\tt RM-CLEAN} component that is $> 50$\% of the $|F(\phi)|$ amplitude of the primary peak (i.e., two comparable, blended peaks), in our RM grids. Radio sources matching this criteria still have $\phi = \mathrm{RM}$ but we update their RM uncertainty to be more conservative:
\begin{equation}
\delta(\mathrm{RM}_\mathrm{obs,broad}) = M_2\,. \label{eq:rm_err2}
\end{equation}
All radio sources with $m_2 < 0.1$ are considered Faraday simple such that they depict only one FDF peak that is approximately Gaussian and are all included in our RM grids. A catalog of all RM sources used in this work is presented in Appendix \ref{sec:appendix}, which includes sky positions, RM and associated uncertainty, and $m_2$.

\section{Foreground DM and RM estimation} \label{sec:foregrounds}
To derive magnetoionic properties of FRB local environments and host galaxies, we are chiefly interested in obtaining $\mathrm{DM}_{\mathrm{host}}$ and $\mathrm{RM}_{\mathrm{host}}$. Thus, our goal is to precisely account for, and subtract, all other foreground components in Equations \ref{eq:dm_comps} and \ref{eq:rm_comps}. In the following sections, we describe the methodology for estimating each of these foreground components.

\subsection{Galactic and IGM DM contribution} \label{subsec:dm_gal}
To estimate the DM contribution from the FRB environment and host galaxy, $\mathrm{DM}_\mathrm{host}$, we need to estimate and subtract both the Galactic DM component ($\mathrm{DM}_\mathrm{MW}$) and, if possible, the IGM component ($\mathrm{DM}_\mathrm{IGM}$). Encompassed within the $\mathrm{DM}_\mathrm{MW}$ term are both the DM contributions from the MW disk as well as from its halo. While multiple models of the Galactic disk DM contribution exist \citep[e.g.,][]{2002astro.ph..7156C, 2024A&A...690A.314H}, we opt to use the thermal electron density $n_\mathrm{e}$ model of the Galactic disk by \cite{2002astro.ph..7156C} (hereafter NE2001) as it is the most commonly used Galactic DM model for FRB studies and allows for direct comparison of our results to previous literature.
We integrate through the full extent of the MW towards the best fit FRB position with the {\tt PyGEDM} package \citep[V3.1.1;][]{2021PASA...38...38P} to obtain the DM contribution from the MW disk. For the MW halo, we assume a constant DM contribution as a function of sky position with a fiducial value of $30~\mathrm{pc}~\mathrm{cm}^{-3}$ based on estimates by \cite{2015MNRAS.451.4277D}, \cite{2020ApJ...888..105Y}, and \cite{2023ApJ...946...58C}. Further, if the FRB has been associated with a host galaxy that has a measured redshift, then the average $\mathrm{DM}_{\mathrm{IGM}}(z)$ along that LOS can be calculated following the linear $\mathrm{DM}_{\mathrm{IGM}}-z$ relation \citep[i.e., the ``Macquart relation'';][]{2020Natur.581..391M}. In this work, we use the $\mathrm{DM}_{\mathrm{IGM}}-z$ relation and its associated uncertainties as presented by \cite{2024ApJ...965...57B}.

\subsection{Galactic RM contribution} \label{subsec:rm_gal}
The interpolation technique to constrain the Galactic RM component used in this work is based on Information Field Theory \citep[IFT;][]{niftyre}, a Bayesian statistical framework to design non-parametric inference algorithms. The specialized application of an IFT algorithm for RM grids was first applied to all sky data sets by \cite{2012A&A...542A..93O, 2015A&A...575A.118O}, \cite{2020A&A...633A.150H}, and recently H22. It has also been used to constrain the Galactic DM and LOS magnetic field skies \citep{2022MNRAS.516.4739P, 2024A&A...690A.314H}. For discrete sky patches, the RM interpolation method was tested by \citet{affan} and we refer the reader to this work for in-depth discussions of performance, possible biases, and comparisons to other techniques. Compared to the setup of \cite{affan}, we have simplified the algorithm in that we only infer a single (Gaussian) field, instead of a product of an amplitude and sign field. This is owing to the relatively sparse sky sampling in some fields (i.e., $\sim 9$ polarized sources per square degree compared to the $\sim 45$ polarized sources per square degree used by \cite{affan}, which makes a reduction of the parameter space necessary to stabilize the inference algorithm and avoid overfitting. The price for that is a reduced capability to represent strong variations (i.e., several orders of magnitude) in RM amplitude within a patch. Our model together with our choice of hyperparameters results in a marginal per pixel prior that allows RM amplitudes well over $\sim 100~\mathrm{rad}~\mathrm{m}^{-2}$. Since we are only reconstructing the $\mathrm{RM}_\mathrm{MW}$ in small regions of the sky around FRBs at high Galactic latitude ($|b| > 25~\mathrm{deg}$), this range easily encompasses the expected amount of RM amplitude variations across our reconstruction area. Thus, we expect no meaningful impact of this modeling choice on the reconstructed $\mathrm{RM}_\mathrm{MW}$ maps.

Another significant difference to the \citet{affan} implementation is that we included a noise estimation feature in the inference algorithm that automatically detects and down-weights extreme outliers in the data. This is useful for removing RM grid sources that are dominated by RM contributions local to the emitting source, and therefore are less suitable tracers of $\mathrm{RM}_\mathrm{MW}$. For details of this feature we refer the reader to \citet{2020A&A...633A.150H}. The reliability of this outlier removal feature critically relies on number statistics, i.e., the more data points constraining the MW component, the more likely we are able to correctly down-weight data points with a strong source contribution.
For this reason, we have used RM data points from the catalog provided by \citet{2023ApJS..267...28V} in addition to the VLA observations when reconstrucing maps of $\mathrm{RM}_\mathrm{MW}$. Given the relatively sparse sky sampling of these additional RMs, these mostly help to constrain the large angular scales (with respect to the field of view) around each FRB. We can test the impact of the VLA RMs by comparing our results to the all sky results of H22, which is constrained by the same catalog (see Section \ref{subsec:rm_map_results}). We note that the additional RMs are largely of poorer quality compared to our VLA grid, as the catalog is dominated by RMs from the NVSS survey \citep{2009ApJ...702.1230T}, which were inferred from only two frequency bands. Possible problems resulting from this were discussed in detail by \citet{2019MNRAS.487.3454M}, but \citet{2020A&A...633A.150H} demonstrated that the very same noise estimation feature as discussed above was able to reliably find and filter these problematic RMs. We have used sources from \cite{2023ApJS..267...28V} within a radius of three times the maximum extent of the respective VLA RM grid (i.e., within a $1.5$~deg radius). We chose this limit, as sources further out will likely not be informative for the scales of the respective VLA field of view.  

\subsection{Ionospheric RM contribution} \label{rm_iono}
The ionospheric RM contribution depends on the LOS through the ionosphere, but also undergoes temporal variations related to the time of day and solar activity, so we need to measure $\mathrm{RM}_{\mathrm{ion}}$ at the sky position and at time of arrival of each FRB. First, a magnetoionic model for the ionosphere is created by combining maps of the vertical total electron content in the ionosphere recorded using a distribution of Global Positioning System satellites with models of the Earth's magnetic field at the time of arrival of an FRB. Then, an estimate of $\mathrm{RM}_\mathrm{ion}$ is computed by integrating over the estimated trajectory of the FRB emission through the ionosphere. For this $\mathrm{RM}_\mathrm{ion}$ estimation, we make use of the {\tt RMextract} package of \cite{2018ascl.soft06024M}.

\section{Results} \label{sec:results}
\setlength{\tabcolsep}{3pt}
\setlength{\LTcapwidth}{1.0\textwidth}
\begin{table*}[ht!]
\begin{center}
\caption{Summary of RM grid statistics for all eight VLA fields.} \label{tb:rm_grids}
\begin{tabular}{ccccccc}
\hline
\hline
FRB Source & \# of observed & \# of RM grid & RM grid density & Median $\mathrm{RM}_\mathrm{obs}$ & Median $\delta(\mathrm{RM}_\mathrm{obs})$ & $\mathrm{RM}_\mathrm{obs}$ MADFM\\
 & sources & sources &  &  &  & standard deviation\\
 &  &  & (deg$^{-2}$) & (rad~m$^{-2}$) & (rad~m$^{-2}$) & (rad~m$^{-2}$)\\
\hline
FRB 20110523A & $35$ & $7$ & $9$ & $13.7$ & $7.3$ & $10.5$\\
FRB 20160102A & $25$ & $15$ & $19$ & $23.1$ & $6.4$ & $16.9$\\
FRB 20181030A & $24$ & $8$ & $10$ & $-23.2$ & $17.7$ & $12.0$\\
FRB 20190117A & $24$ & $16$ & $20$ & $-38.5$ & $9.1$ & $16.7$\\
FRB 20190213B & $19$ & $12$ & $15$ & $10.7$ & $7.4$ & $18.8$\\
FRB 20190608B & $17$ & $10$ & $13$ & $-22.7$ & $8.2$ & $15.3$\\
FRB 20191108A & $18$ & $9$ & $11$ & $-64.1$ & $10.7$ & $32.7$\\
FRB 20200120E & $37$ & $22$ & $28$ & $-17.1$ & $5.6$ & $21.0$\\
\hline
\hline
\end{tabular}
\end{center}
\end{table*}

\subsection{RM grids} \label{subsec:rm_grid_results}
Following the steps laid out in Section \ref{subsubsec:rm_grids}, we construct RM grids for each of our eight VLA fields. In total, only $4$ sources were omitted from our RM grids due to Faraday complexity ($m_2 > 0.25$). In Table \ref{tb:rm_grids} we describe the statistics for each RM grid including the number of observed sources in the field, number of RM grid sources per field, RM grid density, median $\mathrm{RM}_\mathrm{obs}$, median RM measurement uncertainty $\delta(\mathrm{RM}_\mathrm{obs})$, and a robust standard deviation of the $\mathrm{RM}_\mathrm{obs}$ distribution calculated using the median absolute deviation from the median (MADFM) which approximates to $\sim 1.4826 \times \mathrm{MADFM}$. Our RM grid density ranges between $9$ and $28$ sources per square degree. Note that a larger fraction of the observed radio sources in the FRB 20110523A field are unpolarized compared to the other seven fields because the 2018 VLA observing campaign selected fainter targets and spent less time observing each target compared to the 2023 observations. As a result, our RM grid density of the FRB 20110523A field is relatively low ($9~\mathrm{sources}~\mathrm{deg}^{-2}$). All eight fields have a moderate median $\mathrm{RM}_\mathrm{obs}$ magnitude on the order of $10~\mathrm{rad}~\mathrm{m}^{-2}$, as is generally expected for regions away from the Galactic plane \citep[e.g., see][]{2024AJ....167..226V}.

\subsection{Reconstructed RM\texorpdfstring{$_\mathrm{MW}$}{MW} maps} \label{subsec:rm_map_results}
\begin{figure*}
    \centering
    \includegraphics[width=0.893\textwidth]{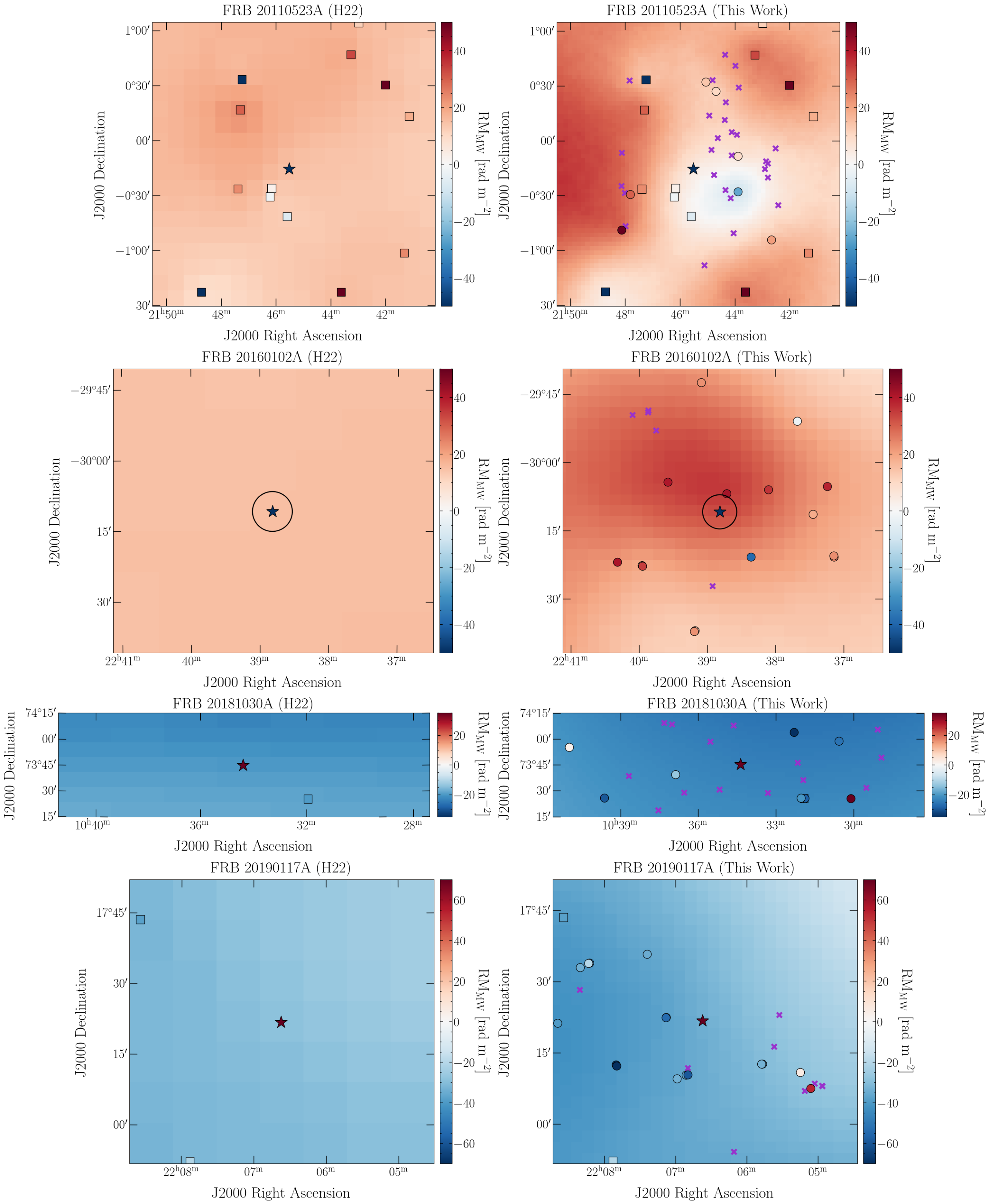}
    \caption{Comparison of the H22 $\mathrm{RM}_\mathrm{MW}$ maps (left) and those reconstructed in this work (right) for FRBs 20110523A, 20160102A, 20181030A, and 20190117A, in order from top to bottom. Each FRB is presented by a star with a black ellipse representing its position uncertanity (in most cases the position uncertainty is smaller than the symbol). RMs of polarized radio galaxies observed prior to this work are shown as squares in both panels and our VLA RM grid sources (Faraday simple sources or those with $0.1 < m_2 < 0.25$) are shown as circles in the right panels. Sources that were observed but did not meet our linear polarization threshold of $(\mathrm{S/N})_P > 8$ are plotted as purple crosses in the right panels. The few Faraday complex sources with $m_2 > 0.25$ are not plotted. The background maps, FRBs, and polarized radio galaxies are all colored by their $\mathrm{RM}_\mathrm{obs}$. The aspect ratio of each panel depends on the Declination of the field (i.e., more compact closer to $\mathrm{Declination} = \pm 90~\mathrm{degrees}$).} 
    \label{fig:rm_maps_1}
\end{figure*}

\begin{figure*}
    \centering
    \includegraphics[width=0.893\textwidth]{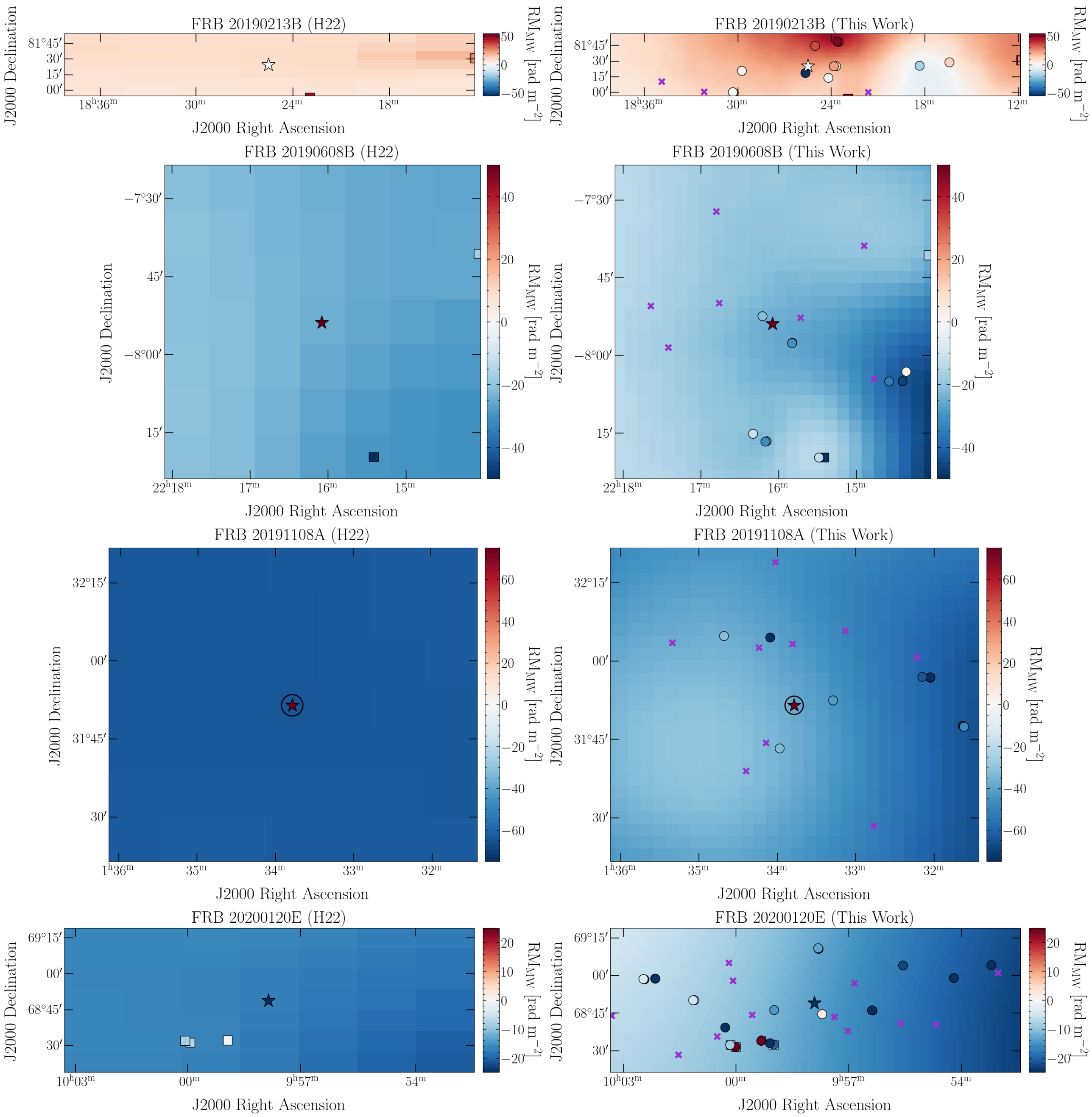}
    \caption{As for Figure \ref{fig:rm_maps_1} but for FRBs 20190213B, 20190608B, 20191108A, and 20200120E.} 
    \label{fig:rm_maps_2}
\end{figure*}

\begin{figure*}
    \centering
    \includegraphics[width=0.893\textwidth]{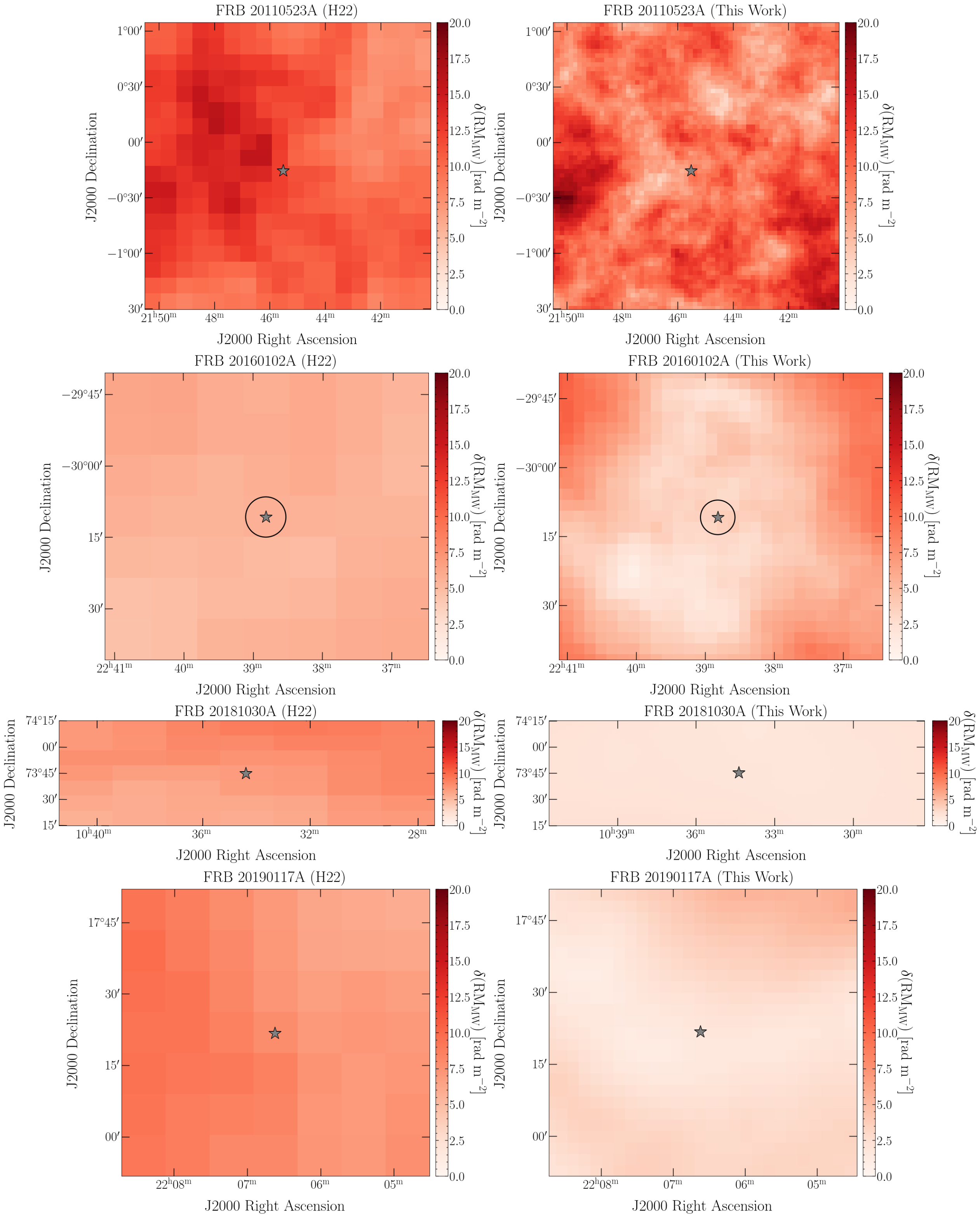}
    \caption{Comparison of the $\mathrm{RM}_\mathrm{MW}$ uncertainty maps from H22 (left) and from this work (right) for FRBs 20110523A, 20160102A, 20181030A, and 20190117A (in order from top to bottom). The position of each FRB is plotted as grey star and their position uncertainty is plotted as black ellipse. In most cases, the position uncertainty is smaller than the symbol.} 
    \label{fig:rm_err_maps_1}
\end{figure*}

\begin{figure*}
    \centering
    \includegraphics[width=0.893\textwidth]{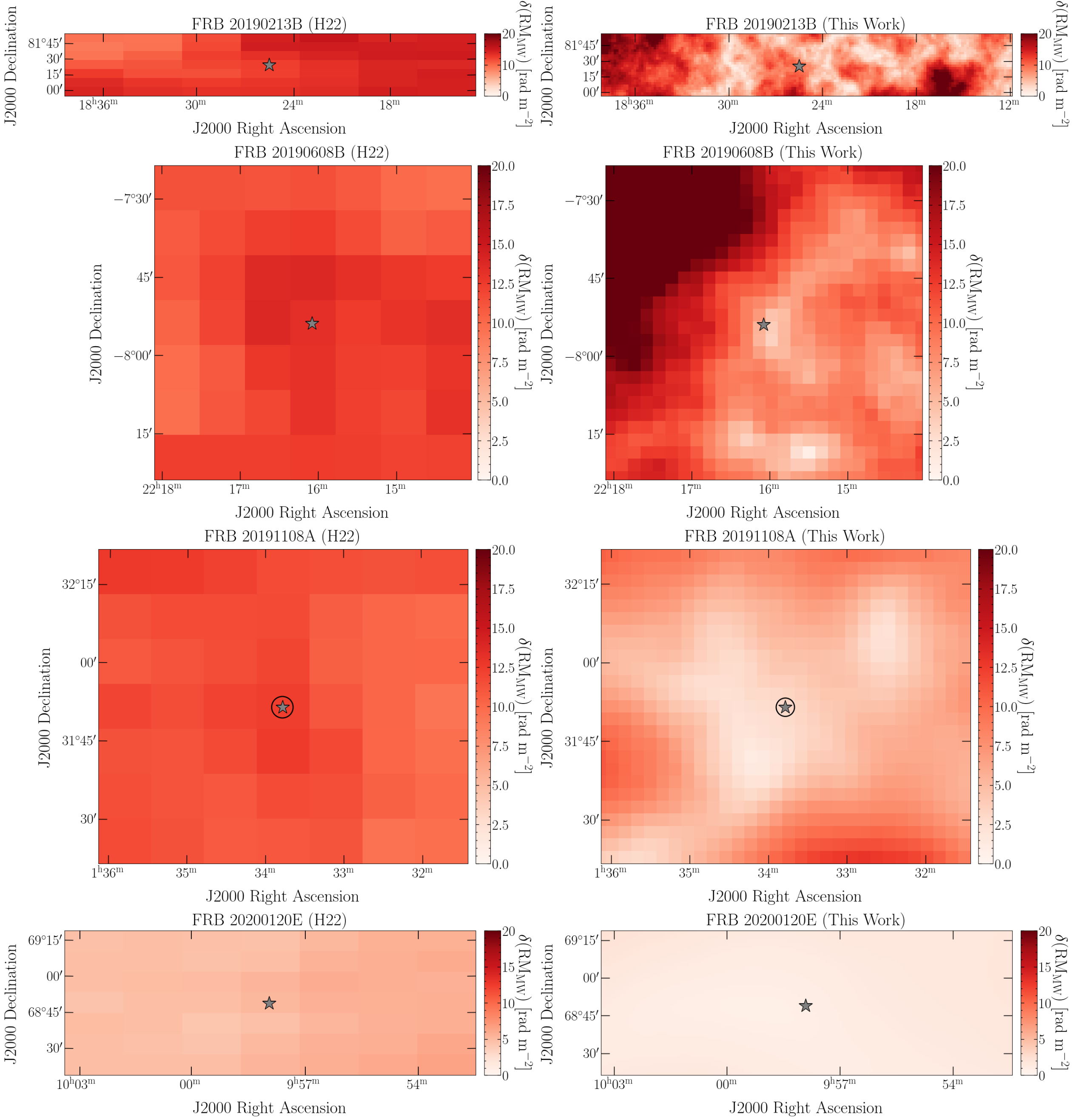}
    \caption{As for Figure \ref{fig:rm_err_maps_1} but for FRBs 20190213B, 20190608B, 20191108A, and 20200120E} 
    \label{fig:rm_err_maps_2}
\end{figure*}

Using the Bayesian inference scheme described in Section \ref{subsec:rm_gal}, we derive reconstructed $\mathrm{RM}_\mathrm{MW}$ and associated statistical uncertainty maps. The reconstructed $\mathrm{RM}_\mathrm{MW}$ maps for each region are presented in the right-hand panels of Figures \ref{fig:rm_maps_1}$-$\ref{fig:rm_maps_2}. For comparison, the H22 $\mathrm{RM}_\mathrm{MW}$ maps are shown in the left-hand panels. The pixel scale for the H22 maps in a cartesian projection is $0.1435~\mathrm{deg}$ and for the reconstructed $\mathrm{RM}_\mathrm{MW}$ maps in this work it is $0.04~\mathrm{deg}$. In the center of each map, we show the respective FRB sky position as a star (with a black ellipse indicating its positional uncertainty) that is coloured by its $\mathrm{RM}_\mathrm{obs}$. In both the left- and right-hand panels, we overplot the previously existing extragalactic RM data in these regions from \cite{2023ApJS..267...28V} as squares. In the right-hand panels, we also plot the VLA RM grids derived in this work as circles and the unpolarized radio sources are shown as purple crosses. The aspect ratio of the plots depend on the Declination of the field (i.e., the vertical axis becomes more compact closer to $\mathrm{Declination} = \pm 90~\mathrm{degrees}$). Notably, the FRB 20110523A field has a RM grid density $> 1$ source per square degree even prior to the inclusion of our VLA observations. This is because that region of the sky overlaps with the first data release of the Spectral and Polarisation in Cutouts of Extragalactic sources from The Rapid Australian Square Kilometer Array Pathfinder Continuum Survey (SPICE-RACS), which provides an average RM grid density of $\sim 4$ sources per square degree \citep{2023PASA...40...40T}. In Figures \ref{fig:rm_err_maps_1}$-$\ref{fig:rm_err_maps_2}, we plot the associated RM statistical uncertainty maps from the Bayesian inference. Again, the H22 maps are on the left and those from our reconstruction are on the right and the FRB position, and associated positional uncertainty, are plotted as a star and black ellipse, respectively. We note that the error maps, in parts, have different degrees of smoothness than the $\mathrm{RM}_\mathrm{MW}$ maps. This reflects the very heterogeneous structure of the RM grids, both with respect to spatial structure as well as measurement uncertainty on the individual $\mathrm{RM}_\mathrm{obs}$ measurements. This leads to varying degrees of certainty of our reconstruction within a given field, even if the posterior mean appears to be smoothly varying. In Figures \ref{fig:rm_maps_1}$-$\ref{fig:rm_maps_2}, six out of eight of our reconstructed $\mathrm{RM}_\mathrm{MW}$ maps show RM variations on $\lesssim 1~\mathrm{deg}^{2}$ scales that were previously unresolved by the H22 all-sky reconstruction, or provide a more granular picture of larger scale RM variations. Only the field around FRBs 20181030A does not show any clear RM variations across the map, while around FRB 20190117A we recover a similar large scale RM gradient as seen in the H22 map. Even in fields without drastic small-scale RM variations, our reconstruction is still notably different from the H22 map, suggesting that we are also able to better constrain the large-scale RM variations with the addition of our localized, high-density RM grids. Qualitatively, in five of eight fields (FRBs 20110523A, 20160102A, 20190213B, 20190608B, and 20191108A), we uncover sharp small-scale variations in $\mathrm{RM}_\mathrm{MW}$ (approximately $\sim 20-40~\mathrm{rad}~\mathrm{m}^{-2}$ variations over $\sim 0.3 - 1~\mathrm{deg}$ scales). In the FRB 20200120E field we discover a previously unseen large-scale RM gradient that shows a change of $\sim 30~\mathrm{rad}~\mathrm{m}^{-2}$ over $\sim 3~\mathrm{deg}$. In all of these cases, the range of $\mathrm{RM}_\mathrm{MW}$ values in our reconstruction is broader than the H22 map. For example, the H22 map shows a constant $\sim 15~\mathrm{rad}~\mathrm{m}^{-2}$ across the FRB 20160102A field, while the $\mathrm{RM}_\mathrm{MW}$ values in our reconstruction range over $\sim 9 - 35~\mathrm{rad}~\mathrm{m}^{-2}$. In two maps (corresponding to FRBs 20110523A and 20190213B), we find changes in the RM sign that potentially suggest a flip in the average LOS magnetic field orientation across the sky. We caution that in both of these fields, the RM sign flips appear to be driven largely by one RM source and, therefore, could instead be RM contribution intrinsic to the source. Furthermore, the negative RM regions in these two maps have $\mathrm{RM}_\mathrm{MW}$ amplitudes close to $0~\mathrm{rad}~\mathrm{m}^{-2}$, which are comparable to or smaller than the $\mathrm{RM}_\mathrm{MW}$ uncertainties in these areas, meaning that they may still be consistent with there being no physical RM sign flips.

Furthermore, as is clear in Figures \ref{fig:rm_err_maps_1}$-$\ref{fig:rm_err_maps_2}, we reduce the average statistical uncertainty on $\mathrm{RM}_\mathrm{MW}$ in all eight regions. The second panel of Figure \ref{fig:rm_diff} also shows that the $\mathrm{RM}_\mathrm{MW}$ uncertainty at the position of all eight FRBs is lower in our reconstruction than in H22. Note that in the case of FRB 20190608B, the statistical $\mathrm{RM}_\mathrm{MW}$ uncertainty in the north-eastern area is not constrained due to a lack of RM data points in that part of the sky. The reason that this issue only manifests in our reconstruction, and not in the H22 map, is because our inference model is computed only over a small region rather than over the entire sky.

\subsection{FRB magnetoionic environments} \label{frb_env_results}
\begin{figure}
    \centering
    \includegraphics[width=0.475\textwidth]{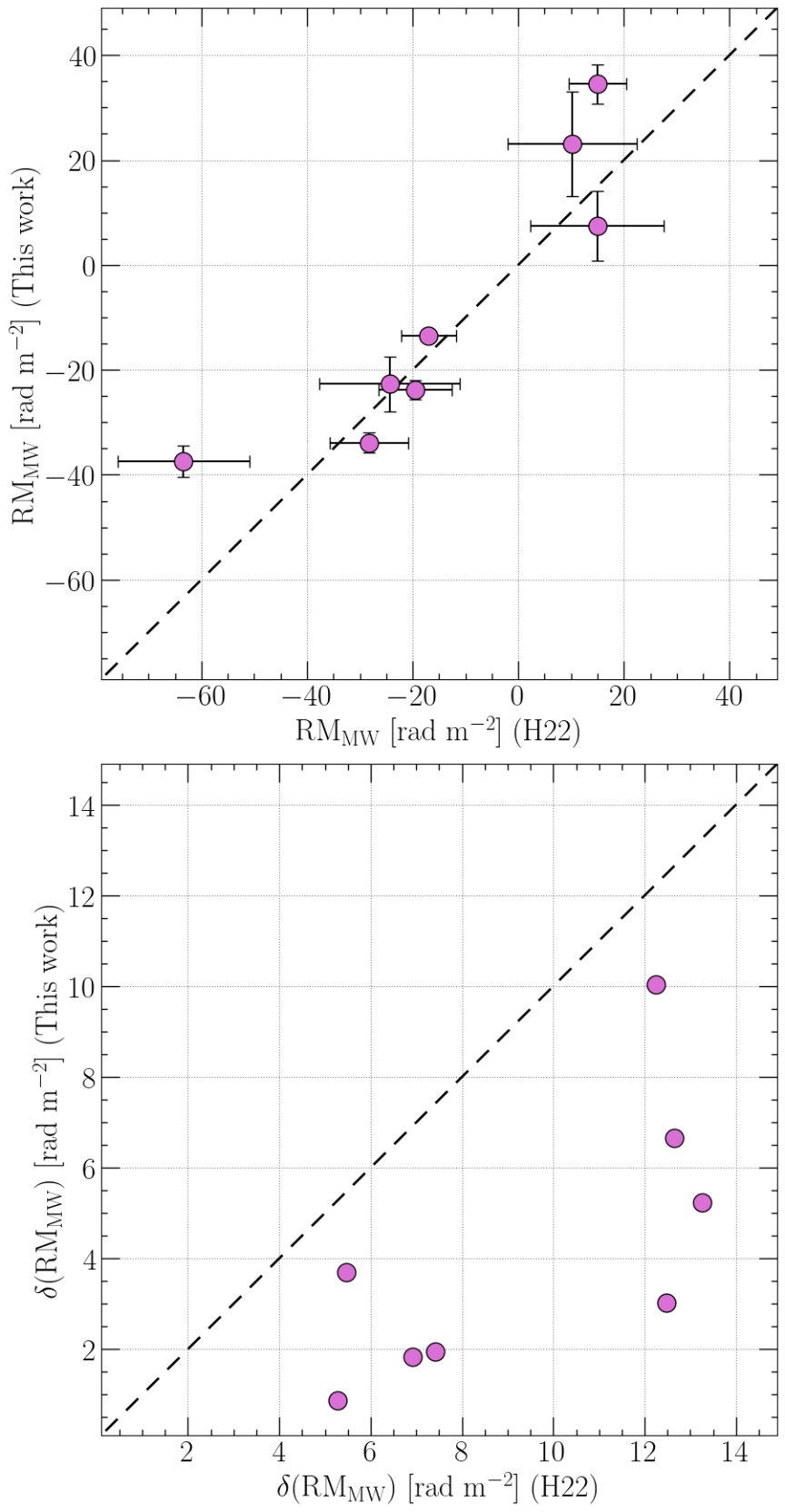}
    \caption{The $\mathrm{RM}_\mathrm{MW}$ (top) and $\delta(\mathrm{RM}_\mathrm{MW})$ (bottom) estimates at the position of our eight FRBs between the H22 map and our reconstructed map using VLA RM grids. In both panels, the black dashed line corresponds to unity between the respective estimates.} 
    \label{fig:rm_comp}
\end{figure}

In the top panel of Figure \ref{fig:rm_comp}, we compare the $\mathrm{RM}_\mathrm{MW}$ specifically towards each of the eight FRBs as determined by H22 (x-axis) and by our RM reconstruction (y-axis). For five of the eight FRBs, the $\mathrm{RM}_\mathrm{MW}$ between the two reconstructions is consistent within statistical uncertainties. In the other three cases, the $\mathrm{RM}_\mathrm{MW}$ between reconstructions is inconsistent, with the largest difference being $26 \pm 13~\mathrm{rad}~\mathrm{m}^{-2}$ for FRB 20191108A. We compare the statistical uncertainties on $\mathrm{RM}_\mathrm{MW}$ between the two reconstructions in the bottom panel of Figure \ref{fig:rm_comp}. In all cases, the uncertainty is improved by a factor of $2-6$ with our updated $\mathrm{RM}_\mathrm{MW}$ reconstruction.

When redshift information for the FRB host galaxy is known, we derive its $\mathrm{DM}_\mathrm{host}$ and $\mathrm{RM}_\mathrm{host}$ following Equations \ref{eq:dm_comps} and \ref{eq:rm_comps}, respectively. Using these values, we can calculate the electron density-weighted average magnetic field strength parallel to the LOS in the FRB host galaxy as:
\begin{equation}
\left<B_{\parallel,\mathrm{host}}\right> = 1.232 \frac{\mathrm{RM}_\mathrm{host}}{\mathrm{DM}_\mathrm{host}}~\mu\text{G}\,. \label{eq:b_host}
\end{equation}
If the redshift of the FRB host galaxy is unknown, we instead compute the observer frame values:
\begin{equation}
\mathrm{DM}_{\mathrm{EG}} = \mathrm{DM}_{\mathrm{obs}} - \mathrm{DM}_{\mathrm{MW}}~\mathrm{pc}~\mathrm{cm}^{-3}\,; \label{eq:dm_eg}
\end{equation}
\begin{equation}
\mathrm{RM}_{\mathrm{EG}} = \mathrm{RM}_{\mathrm{obs}} - \mathrm{RM}_{\mathrm{MW}}~\mathrm{rad}~\mathrm{m}^{-2}\,. \label{eq:rm_eg}
\end{equation}
Here the subscript ``EG'' denotes the combined extragalactic contribution from the IGM and host galaxy in the observer frame. Then, the observer frame analog of $\left<B_{\parallel,\mathrm{host}}\right>$ is:
\begin{equation}
\beta = 1.232 \frac{\mathrm{RM}_\mathrm{EG}}{\mathrm{DM}_\mathrm{EG}}~\mu\text{G}\,. \label{eq:beta}
\end{equation}

The uncertainties in some of our foreground estimates are strongly non-Gaussian and asymmetric. Below, we list the distributions and associated uncertainties for every foreground estimate in our DM and RM decomposition:
\begin{enumerate}
    \item $\mathrm{DM}_\mathrm{obs}$: Drawn from a Gaussian distribution with mean equal to the measured $\mathrm{DM}_\mathrm{obs}$ and the standard deviation equal to the associated measurement uncertainty. We note here that a Gaussian distribution is not always a reliable assumption for modeling $\mathrm{DM}_\mathrm{obs}$, which is a definite positive quantity, as it could lead to randomly drawn samples with negative $\mathrm{DM}_\mathrm{obs}$. In our case, the combination of relatively large $\mathrm{DM}_\mathrm{obs}$ for our FRBs ($\gtrsim 88~\mathrm{pc}~\mathrm{cm}^{-3}$) and minuscule measurement uncertainties ($\lesssim 0.3~\mathrm{pc}~\mathrm{cm}^{-3}$) means that there is an extremely small likelihood of drawing a negative $\mathrm{DM}_\mathrm{obs}$. Therefore, the choice of a Gaussian distribution for $\mathrm{DM}_\mathrm{obs}$ does not effect our results. However, we caution against the use of this assumption in cases where $\mathrm{DM}_\mathrm{obs}$ is very small and/or the associated measurement uncertainity is large.
    \item $\mathrm{DM}_\mathrm{disk}$: Drawn from a Uniform distribution with 20\% uncertainties about the NE2001 prediction at the given sky position \citep{2002astro.ph..7156C}.
    \item $\mathrm{DM}_\mathrm{halo}$: We assumed 0.2 dex uncertainty about the fiducial $30~\mathrm{pc}~\mathrm{cm}^{-3}$ used in this work \citep[i.e., a Log-Normal distribution with mean $\mathrm{ln}(30)$ and variance $\mathrm{ln}(10^{0.2})$;][]{2020ApJ...888..105Y}.
    \item $\mathrm{DM}_\mathrm{IGM}$: Directly sampled from the probability distribution presented in Equation 1 by \cite{2024ApJ...965...57B}. The $\mathrm{DM}_\mathrm{IGM}$ is only included in the DM decomposition if the FRB has a known redshift.
    \item $\mathrm{RM}_\mathrm{obs}$: Drawn from a Gaussian distribution with mean equal to the measured $\mathrm{RM}_\mathrm{obs}$ and the standard deviation equal to the associated measurement uncertainty.
    \item $\mathrm{RM}_\mathrm{ion}$: We use the estimate obtained from {\tt RMextract} \citep{2018ascl.soft06024M} without an associated uncertainty since $\mathrm{RM}_\mathrm{ion}$ is much smaller than the other RM contributions along the LOS.
    \item $\mathrm{RM}_\mathrm{IGM}$: Drawn from a Gaussian distribution with mean $0~\mathrm{rad}~\mathrm{m}^{-2}$ and standard deviation $6~\mathrm{rad}~\mathrm{m}^{-2}$ following \cite{2010MNRAS.409L..99S}.
    \item $\mathrm{RM}_\mathrm{MW}$: Both estimate and associated uncertainty obtained from either the H22 maps or the reconstructed maps derived in this work.
\end{enumerate}

\begin{figure}
    \centering
    \includegraphics[width=0.475\textwidth]{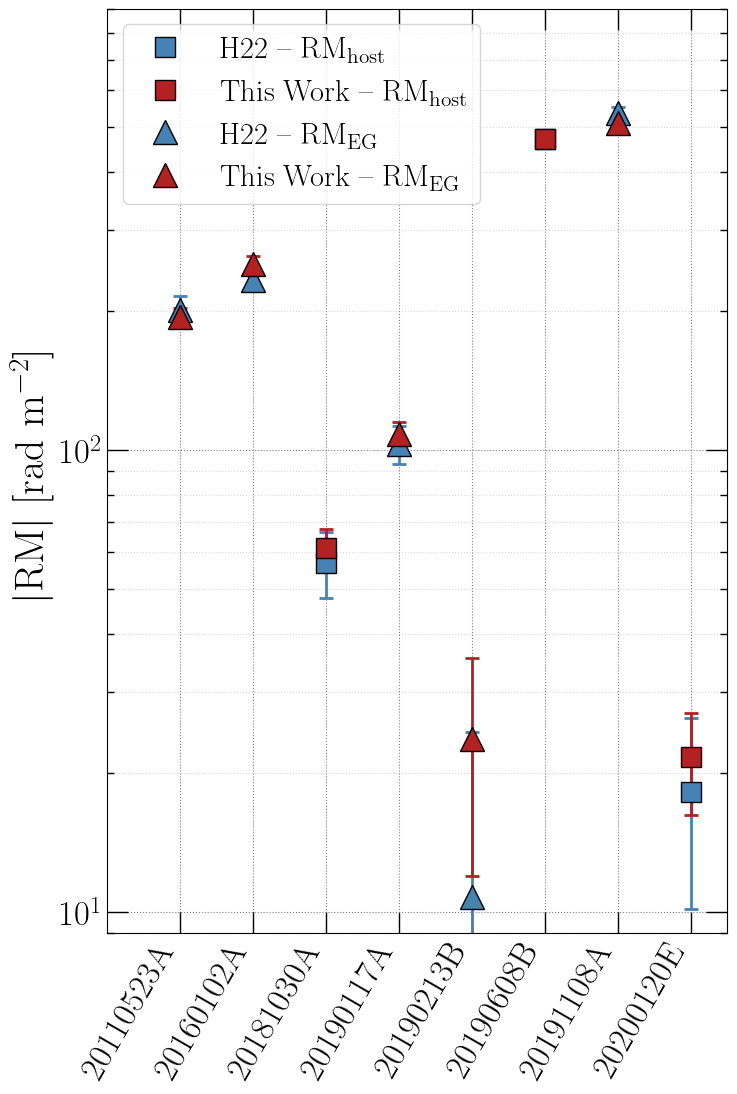}
    \caption{Comparison of the $|\mathrm{RM}_\mathrm{host}|$ (squares) or $|\mathrm{RM}_\mathrm{EG}|$ (triangles) for all eight FRBs as estimated using the H22 $\mathrm{RM}_\mathrm{MW}$ map (blue) versus the reconstructions from this work (red). The y-axis scaling is logarithmic to highlight the fractional change in $|\mathrm{RM}_\mathrm{host}|$).} 
    \label{fig:rm_diff}
\end{figure}

Thus, we compute uncertainties on our estimates of $\mathrm{DM}_\mathrm{host}$, $\mathrm{RM}_\mathrm{host}$, and $\left<B_{\parallel,\mathrm{host}}\right>$ (or $\mathrm{DM}_\mathrm{EG}$, $\mathrm{RM}_\mathrm{EG}$, and $\beta$) using Monte Carlo random sampling. In this procedure, we randomly draw values for all foreground DM and RM components from their known underlying distributions, use those to compute $\mathrm{DM}_\mathrm{host}$, $\mathrm{RM}_\mathrm{host}$, and $\left<B_{\parallel,\mathrm{host}}\right>$ (or $\mathrm{DM}_\mathrm{EG}$, $\mathrm{RM}_\mathrm{EG}$, and $\beta$), and we repeat this process for 10,000 trials. We approximate our uncertainties as the $68$\% confidence interval around the median from the resulting $\mathrm{DM}_\mathrm{host}$, $\mathrm{RM}_\mathrm{host}$, and $\left<B_{\parallel,\mathrm{host}}\right>$ (or $\mathrm{DM}_\mathrm{EG}$, $\mathrm{RM}_\mathrm{EG}$, and $\beta$) distributions. In the second section of Table \ref{tb:frb_decomp}, we tabulate the $\mathrm{DM}_\mathrm{host}$, $\mathrm{RM}_\mathrm{host}$, and $\left<B_{\parallel,\mathrm{host}}\right>$ (or $\mathrm{DM}_\mathrm{EG}$, $\mathrm{RM}_\mathrm{EG}$, and $\beta$) of each FRB, computed using the H22 map and using our updated reconstruction. We quantify the impact of the updated $\mathrm{RM}_\mathrm{MW}$ reconstruction on our understanding of the FRB magnetoionic environments in Figure \ref{fig:rm_diff}. Specifically, we compare the magnitude of $\mathrm{RM}_\mathrm{host}$ (or $\mathrm{RM}_\mathrm{EG}$) predicted by both reconstructions for each FRB.

\setlength{\tabcolsep}{8pt}
\setlength{\LTcapwidth}{1.0\textwidth}
\begin{table*}
\begin{center}
\caption{Summary of DM and RM decomposition for each of the eight FRBs.} \label{tb:frb_decomp}
\begin{tabular}{cccccc}
\hline
\hline
\multicolumn{6}{c}{\textbf{Foreground contributions}}\\
\hline
FRB Source & ${\mathrm{DM}_\mathrm{MW}}$ & ${\mathrm{DM}_\mathrm{IGM}}$ & ${\mathrm{RM}_\mathrm{ion}}$ & ${\mathrm{RM}_\mathrm{MW}}$ (H22) & ${\mathrm{RM}_\mathrm{MW}}$ (This work)\\
 & (pc cm$^{-3}$) & (pc cm$^{-3}$) & (rad m$^{-2}$) & (rad m$^{-2}$) & (rad m$^{-2}$)\\
\hline
FRB 20110523A & $73.4^{+17.6}_{-12.1}$ & $-$ & $2.4$ & $14.9 \pm 12.6$ & $7.5 \pm 6.7$\\
FRB 20160102A & $64.5^{+17.7}_{-11.8}$ & $-$ & $-2.0$ & $14.9 \pm 5.5$ & $34.5 \pm 3.7$\\
\textbf{FRB 20181030A} & $71.1^{+17.4}_{-12.3}$ & $3.2^{+1.7}_{-1.0}$ & $0.24$ & $-19.6 \pm 6.9$ & $-23.8 \pm 1.8$\\
 &  &  & $0.23$ &  & \\
 &  &  & $0.51$ &  & \\
\textbf{FRB 20190117A} & $78.2^{+17.6}_{-12.5}$ & $-$ & $0.84$ & $-28.3 \pm 7.4$ & $-33.9 \pm 1.9$\\
 &  &  & $0.44$ &  & \\
\textbf{FRB 20190213B} & $82.7^{+18.0}_{-13.0}$ & $-$ & $0.11$ & $10.2 \pm 12.2$ & $23.1 \pm 10.0$\\
 &  &  & $0.34$ &  & \\
 &  &  & $0.69$ &  & \\
 &  &  & $0.54$ &  & \\
 &  &  & $0.38$ &  & \\
FRB 20190608B & $67.3^{+17.7}_{-12.2}$ & $99.9^{+46.5}_{-26.6}$ & $-0.52$ & $-24.4 \pm 13.3$ & $-22.7 \pm 5.2$\\
FRB 20191108A & $81.8^{+18.3}_{-12.9}$ & $-$ & $0.23$ & $-63.4 \pm 12.4$ & $-37.5 \pm 3.0$\\
\textbf{FRB 20200120E} & $70.7^{+18.0}_{-12.3}$ & $1.0^{\dagger}$ & $0.14$ & $-17.0 \pm 5.3$ & $-13.6 \pm 0.9$\\
 &  &  & $0.34$ &  & \\
 &  &  & $0.20$ &  & \\
 &  &  & $0.20$ &  & \\
 &  &  & $0.41$ &  & \\
 &  &  & $0.39$ &  & \\
\hline
\multicolumn{6}{c}{\textbf{FRB Magnetoionic Environment}}\\
\hline
FRB Source & $\mathrm{DM}_\mathrm{host}$ & $\mathrm{RM}_\mathrm{host}$ (H22) & $\mathrm{RM}_\mathrm{host}$ (This work) & $B_\mathrm{host}$ (H22) & $B_\mathrm{host}$ (This work)\\
 & (pc cm$^{-3}$) & (rad m$^{-2}$) & (rad m$^{-2}$) & ($\mu$G) & ($\mu$G)\\
\hline
FRB 20110523A & $549.6^{+12.2}_{-18.1}$ & $-203.7 \pm 14.1$ & $-196.2 \pm 8.9$ & $-0.46 \pm 0.03$ & $-0.44 \pm 0.02$\\
FRB 20160102A & $2531^{+12.1}_{-17.9}$ & $-233.6 \pm 10.3$ & $-253.1 \pm 9.5$ & $-0.11 \pm 0.01$ & $-0.12 \pm 0.01$\\
\textbf{FRB 20181030A} & $\mathit{30.4^{+11.7}_{-14.4}}$ & $\mathit{56.2 \pm 9.1}$ & $\mathit{60.4 \pm 6.3}$ & $\mathit{2.3^{+2.1}_{-0.7}}$ & $\mathit{2.5^{+2.2}_{-0.7}}$\\
 & $\mathit{30.8^{+11.6}_{-14.5}}$ & $\mathit{57.5 \pm 9.2}$ & $\mathit{61.7 \pm 6.2}$ & $\mathit{2.3^{+2.2}_{-0.7}}$ & $\mathit{2.5^{+2.2}_{-0.7}}$\\
 & $\mathit{30.6^{+11.7}_{-14.8}}$ & $\mathit{57.0 \pm 9.3}$ & $\mathit{61.5 \pm 6.2}$ & $\mathit{2.3^{+2.2}_{-0.7}}$ & $\mathit{2.5^{+2.4}_{-0.7}}$\\
\textbf{FRB 20190117A} & $314.3^{+12.6}_{-18.7}$ & $101.9 \pm 9.6$ & $107.3 \pm 6.3$ & $0.40^{+0.05}_{-0.04}$ & $0.42^{+0.04}_{-0.03}$\\
 & $317.1^{+12.8}_{-17.8}$ & $104.2 \pm 9.5$ & $109.7 \pm 6.4$ & $0.41 \pm 0.04$ & $0.43^{+0.04}_{-0.03}$\\
\textbf{FRB 20190213B} & $218.0^{+13.2}_{-18.2}$ & $-13.9 \pm 13.7$ & $-26.9 \pm 11.8$ & $-0.08 \pm 0.08$ & $-0.15 \pm 0.07$\\
 & $218.1^{+12.8}_{-17.9}$ & $-9.5 \pm 13.5$ & $-22.6 \pm 11.7$ & $-0.05 \pm 0.08$ & $-0.13 \pm 0.07$\\
 & $218.2^{+12.9}_{-18.0}$ & $-9.5 \pm 13.6$ & $-22.7 \pm 11.6$ & $-0.05 \pm 0.08$ & $-0.13 \pm 0.07$\\
 & $218.2^{+13.2}_{-18.3}$ & $-9.7 \pm 13.7$ & $-22.6 \pm 11.7$ & $-0.06 \pm 0.08$ & $-0.13 \pm 0.07$\\
 & $218.2^{+13.2}_{-17.9}$ & $-10.9 \pm 13.6$ & $-23.8 \pm 11.6$ & $-0.06 \pm 0.08$ & $-0.14 \pm 0.07$\\
FRB 20190608B & $\mathit{219.8^{+28.5}_{-55.1}}$ & $\mathit{471.7 \pm 17.9}$ & $\mathit{470.4 \pm 10.2}$ & $\mathit{2.6^{+0.9}_{-0.3}}$ & $\mathit{2.6^{+0.9}_{-0.3}}$\\
FRB 20191108A & $505.8^{+12.8}_{-18.5}$ & $537.2 \pm 14.1$ & $511.2 \pm 7.3$ & $1.3^{+0.06}_{-0.05}$ & $1.2^{+0.05}_{-0.04}$\\
\textbf{FRB 20200120E} & $\mathit{18.9^{+10.1}_{-10.9}}$ & $\mathit{-12.9 \pm 7.9}$ & $\mathit{-16.3 \pm 6.1}$ & $\mathit{-0.8^{+0.5}_{-1.3}}$ & $\mathit{-1.1^{+0.5}_{-1.5}}$\\
 & $\mathit{18.8^{+10.3}_{-11.1}}$ & $\mathit{-10.2 \pm 8.0}$ & $\mathit{-13.6 \pm 6.0}$ & $\mathit{-0.7^{+0.5}_{-1.2}}$ & $\mathit{-0.9^{+0.5}_{-1.3}}$\\
 & $\mathit{19.0^{+9.9}_{-10.8}}$ & $\mathit{-5.2 \pm 15.3}$ & $\mathit{-8.4 \pm 14.5}$ & $\mathit{-0.3^{+1.1}_{-1.3}}$ & $\mathit{-0.6^{+1.0}_{-1.4}}$\\
 & $\mathit{18.8^{+10.4}_{-10.8}}$ & $\mathit{-40.3 \pm 9.5}$ & $\mathit{-43.7 \pm 7.9}$ & $\mathit{-2.9^{+1.1}_{-3.6}}$ & $\mathit{-2.9^{+1.1}_{-3.9}}$\\
 & $\mathit{19.1^{+10.1}_{-11.0}}$ & $\mathit{-20.6 \pm 9.1}$ & $\mathit{-24.0 \pm 7.3}$ & $\mathit{-1.3^{+0.7}_{-1.9}}$ & $\mathit{-1.6^{+0.7}_{-2.1}}$\\
 & $\mathit{19.1^{+10.1}_{-11.0}}$ & $\mathit{-20.1 \pm 11.2}$ & $\mathit{-23.5 \pm 10.1}$ & $\mathit{-1.3^{+0.8}_{-2.0}}$ & $\mathit{-1.5^{+0.8}_{-2.2}}$\\
\hline
\hline
\multicolumn{6}{l}{$^{\dagger}$ $\mathrm{DM}_\mathrm{IGM}$ estimated by \cite{2022Natur.602..585K}; we assume no uncertainties on $\mathrm{DM}_\mathrm{IGM}$ in this specific instance.} \\
\multicolumn{6}{l}{Sources with TNS names presented in bold face are repeating FRBs. For sources with redshift information, we present} \\
\multicolumn{6}{l}{the respective rest frame quantity in italics, otherwise the observer frame quantity is provided. All reported uncertainties} \\
\multicolumn{6}{l}{are the 16$^\mathrm{th}$ and 84$^\mathrm{th}$ percentiles except for $\mathrm{DM}_\mathrm{MW}$, for which they are $20$\% of the NE2001 prediction \citep{2002astro.ph..7156C}.} \\
\multicolumn{6}{l}{} \\
\end{tabular}
\end{center}
\end{table*}

\subsection{The extragalactic RM contribution at 3~GHz} \label{rm_igm_results}
Previous RM surveys have been conducted primarily at or below $1.4$~GHz \citep[e.g.,][]{2009ApJ...702.1230T, 2023MNRAS.519.5723O}. In these studies, a model of the Galactic RM is subtracted from the RMs of the polarized radio galaxies themselves to obtain the sum of the extragalactic RM contributions along that LOS, $\mathrm{RM}_\mathrm{EG}$. We expect to see non-zero variance in the $\mathrm{RM}_\mathrm{EG}$ distribution arising from differing intrinsic source environments, path lengths through the IGM, and intervening structures. At $144$~MHz, $\sigma(\mathrm{RM}_\mathrm{EG})$ has been measured to be $1.8 - 4.0~\mathrm{rad}~\mathrm{m}^{-2}$ \citep{2023MNRAS.519.5723O}, while at $1.4$~GHz it is $6-7~\mathrm{rad}~\mathrm{m}^{-2}$ \citep{2010MNRAS.409L..99S,2015A&A...575A.118O}. The discrepancy between $\sigma(\mathrm{RM}_\mathrm{EG})$ estimates at these two frequencies may be due to stronger depolarization effects at lower frequencies. This hypothesis suggests that higher frequency polarization observations should, on average, probe more complex sight-lines (e.g., with more turbulent intrinsic environments or sight-lines passing through turbulent intervening structures). Here, we derive an estimate of $\sigma(\mathrm{RM}_\mathrm{EG})$ at $3.0$~GHz and compare it to the results at lower frequencies to determine if, on average, our higher frequency data probes more complex sight-lines.

\begin{figure}
    \centering
    \includegraphics[width=0.475\textwidth]{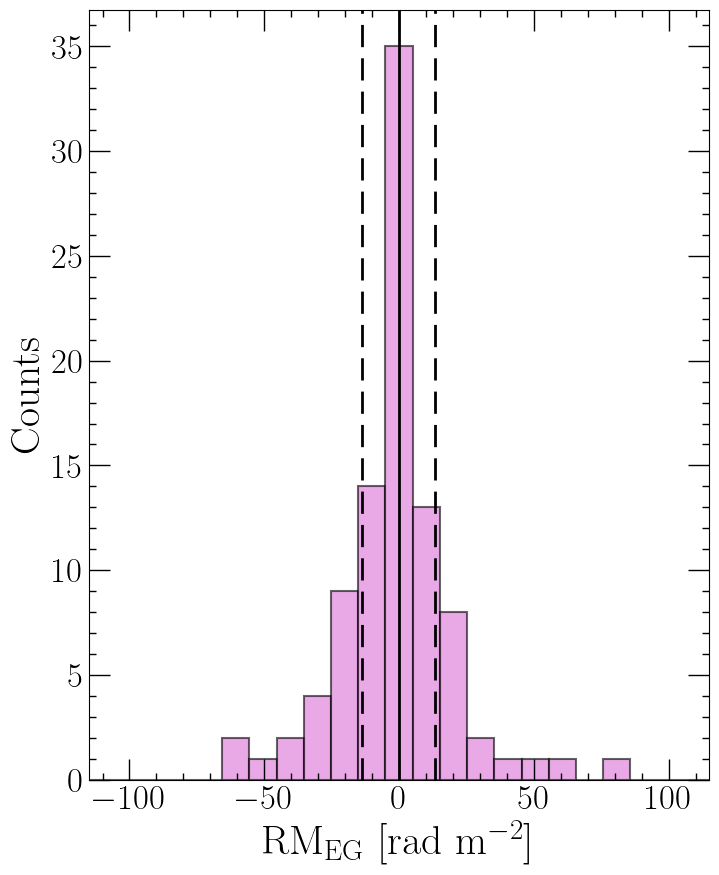}
    \caption{Histogram of the $\mathrm{RM}_\mathrm{EG}$ distribution for the $79$ radio galaxies with $m_2 < 0.1$ and $(\mathrm{S/N})_P > 12$. The median is indicated by a black solid line and the robust standard deviation from the MADFM, $\sigma(\mathrm{RM}_\mathrm{EG,obs})$, is shown as black dashed lines.} 
    \label{fig:rrm_hist}
\end{figure}

To derive an estimate of $\sigma(\mathrm{RM}_\mathrm{EG})$ at $3~\mathrm{GHz}$, we first define a subset of $79$ radio galaxies from our combined RM grid data that are Faraday simple ($m_2 < 0.1$) and that have high linearly polarized signal-to-noise ($(\mathrm{S/N})_P > 12$). For each radio galaxy we subtract the Galactic RM towards that position from our $\mathrm{RM}_\mathrm{MW}$ reconstructions and, in Figure \ref{fig:rrm_hist}, we plot the distribution of $\mathrm{RM}_\mathrm{EG}$ for the $79$ radio galaxies. We overlay the median $\mathrm{RM}_\mathrm{EG}$ ($-0.2~\mathrm{rad}~\mathrm{m}^{-2}$) as a solid black line, and the robust standard deviation calculated using the MAD, $\sigma(\mathrm{RM}_\mathrm{EG,obs}) = 13.6~\mathrm{rad}~\mathrm{m}^{-2}$, is indicated by black dashed lines. Here, we include the subscript ``obs'' as this estimate still includes the variance introduced by various RM uncertainties. We then obtain $\sigma(\mathrm{RM}_\mathrm{EG})$ by subtracting the variance contributed by the RM measurement uncertainty, $\delta(\mathrm{RM}_\mathrm{obs})$, and the statistical error in the $\mathrm{RM}_\mathrm{MW}$ reconstruction, $\delta(\mathrm{RM}_\mathrm{MW})$, following Equation 5 of \cite{2024arXiv240807178O}:
\begin{equation}
\sigma(\mathrm{RM}_\mathrm{EG}) = \left(\sigma^2(\mathrm{RM}_\mathrm{EG,obs}) - \frac{\sum_i^N\left(\delta(\mathrm{RM}_\mathrm{EG})\right)_i^2}{N-1} \right)^{1/2}\,, \label{eq:sigma_rm}
\end{equation}
where $\delta(\mathrm{RM}_\mathrm{EG})$ is the quadrature sum of $\delta(\mathrm{RM}_\mathrm{obs})$ and $\delta(\mathrm{RM}_\mathrm{MW})$, and $N=79$. Following Equation \ref{eq:sigma_rm}, we derive an estimate of $\sigma(\mathrm{RM}_\mathrm{EG}) = 6.8~\mathrm{rad}~\mathrm{m}^{-2}$ at $3.0$~GHz. To quantify an uncertainty on this estimate, we bootstrap (with replacement) over $10,000$ iteration from our $\sigma(\mathrm{RM}_\mathrm{EG,obs})$, $\delta(\mathrm{RM}_\mathrm{obs})$, and $\delta(\mathrm{RM}_\mathrm{MW})$ distributions, and estimate a $\sigma(\mathrm{RM}_\mathrm{EG})$ for each iteration. We adopt the standard deviation of the resampled $\sigma(\mathrm{RM}_\mathrm{EG})$ distribution as the uncertainty on our estimate, so that $\sigma(\mathrm{RM}_\mathrm{EG}) = 6.8 \pm 3.1~\mathrm{rad}~\mathrm{m}^{-2}$. We find that our estimate of $\sigma(\mathrm{RM}_\mathrm{EG})$ at $3.0~\mathrm{GHz}$ is consistent with those derived at $1.4~\mathrm{GHz}$ by \cite{2010MNRAS.409L..99S} and \cite{2015A&A...575A.118O}. We do not find clear frequency dependent variance between $\sigma(\mathrm{RM}_\mathrm{EG})$ measurements at $1.4$ and $3.0$~GHz, but do find that $\sigma(\mathrm{RM}_\mathrm{EG})$ as measured at $144$~MHz is smaller than our estimate. This suggests that while the much lower frequency observations at $144$~MHz are sensitive to depolarization effects and trace LOS that are preferentially less complex, both L-band and S-band observations likely trace similarly complex sight-lines. However, we note that the $144$~MHz observations preferentially probe low $\mathrm{RM}_\mathrm{obs}$ sources due to increased intra-channel depolarization at lower observing frequencies and, therefore, may underestimate $\sigma(\mathrm{RM}_\mathrm{EG})$. Our $\sigma(\mathrm{RM}_\mathrm{EG})$ estimate provides a good reference for future large-sky polarization surveys at $>1.4~\mathrm{GHz}$, such as VLASS \citep{2014arXiv1401.1875M, 2020PASP..132c5001L}.

\section{Discussion} \label{sec:discussion}
\subsection{FRB magnetoionic environments} \label{sec:frb_env_discussion}
By isolating the host galaxy contribution to DM and RM, we can both put constraints on the sources and progenitors of FRBs and identify the subset of FRBs with clean environments that will be the best astrophysical tools for studying intervening media.

We expect different levels of typical $\left| \mathrm{RM} \right|$ contributions from different FRB local environments. On the lower density and weakly magnetized end of the spectrum (e.g., globular clusters), we might expect the local environment to contribute $\sim 0~\mathrm{rad}~\mathrm{m}^{-2}$. For FRBs in the vicinity of star forming regions in a MW-like galaxy, \cite{2019MNRAS.488.4220H} predict a median $\left| \mathrm{RM} \right|$ contribution on the order of $10~\mathrm{rad}~\mathrm{m}^{-2}$ and up to a few $10^{2}~\mathrm{rad}~\mathrm{m}^{-2}$. If instead the FRB is enshrouded by a compact nebula (e.g., a young supernova remnant), we might expect it to contribute up to $\sim 10^3~\mathrm{rad}~\mathrm{m}^{-2}$ \citep{2018ApJ...861..150P}. At high Galactic latitudes ($|b| > 25~\mathrm{deg}$), we find that the $\mathrm{RM}_\mathrm{MW}$ estimate between H22 and our results changes as little as a few $\mathrm{rad}~\mathrm{m}^{-2}$ to $26 \pm 13~\mathrm{rad}~\mathrm{m}^{-2}$. Therefore, our interpretation of the local magnetoionic environment of FRBs with $|\mathrm{RM}_\mathrm{EG}| \lesssim 26~\mathrm{rad}~\mathrm{m}^{-2}$ (assuming $\mathrm{RM}_\mathrm{MW}$ from H22) can change when using higher density RM grids. Based on the $|\mathrm{RM}_\mathrm{EG}|$ distribution of the FRB population studied by \cite{2024ApJ...968...50P}, which is the largest sample of FRB sources with polarization properties to date, this applies to $\sim 30$\% of FRBs. Note, however, that $\sim 30$\% is likely a lower limit, as FRBs at $|b| < 25~\mathrm{deg}$ will, on average, encounter larger magnitude RM variations over similar or smaller scales.

All eight of our FRBs have $\mathrm{RM}_\mathrm{EG}$ or $\mathrm{RM}_\mathrm{host}$ values that are generally in line with various environments in a MW-like galaxy, \cite{2019MNRAS.488.4220H}. Note that those FRBs with $\mathrm{RM}_\mathrm{EG}$ in the observer frame could be at notably larger $\mathrm{RM}_\mathrm{host}$ if the associated host galaxy is at a high redshift. In comparison, the median of the Galactic pulsar $|\mathrm{RM}|$ distribution is approximately $\sim 70~\mathrm{rad}~\mathrm{m}^{-2}$, though the standard deviation is quite large.\footnote{Based on the $1494$ pulsars with RM measurements in the Australia Telescope National Facility pulsar catalog V1.71 \citep{2005AJ....129.1993M}, which is hosted at \href{https://www.atnf.csiro.au/research/pulsar/psrcat/}{https://www.atnf.csiro.au/research/pulsar/psrcat/}.} The variation observed in our sample of eight FRBs, with $\left|\mathrm{RM}_\mathrm{EG}\right|$ or $\left|\mathrm{RM}_\mathrm{host}\right|$ amplitudes spanning $\sim 10 - 500~\mathrm{rad}~\mathrm{m}^{-2}$, could imply variety in their local environments (e.g., in the disk versus the halo, or differing vicinity to active sites of star formation and \ion{H}{2} regions) or distinct positions within their host galaxies (e.g., smaller or larger offset from the center of the host galaxy or different heights with respect to the host galaxy disk). Two sources in particular, FRB 20190608B ($\left|\mathrm{RM}_\mathrm{host}\right| = 470.0~\mathrm{rad}~\mathrm{m}^{-2}$) and 20191108A ($\left|\mathrm{RM}_\mathrm{EG}\right| = 511.2~\mathrm{rad}~\mathrm{m}^{-2}$) are also consistent with predictions from dense circumburst environments \citep[e.g., within a supernova remnant][]{2018ApJ...861..150P}. Since our sample size is small, we do not draw any conclusions about relative magnetoionic environments of repeaters compared to non-repeaters.

We do not find any significant changes to any of the FRB $\beta$ or $B_\mathrm{host}$ estimates after updating the Galactic RM estimate from H22 to our reconstruction. Also, all of our FRBs have $\beta$ or $B_\mathrm{host}$ that is generally consistent with the $\sim \mu\mathrm{G}$ level magnetic fields expected in spiral galaxies \citep{2001SSRv...99..243B}. The five FRBs without known redshifts potentially have a significant (unaccounted for) $\mathrm{DM}_\mathrm{IGM}$ contribution, which would cause $\beta$ to underestimate the rest frame $B_\mathrm{host}$ by a substantial margin. In addition, the uncertainties in the DM decomposition, particularly for $\mathrm{DM}_\mathrm{MW}$ and $\mathrm{DM}_\mathrm{IGM}$, are comparatively larger than those in the RM decomposition. Therefore, we find that improving our understanding of the small-scale Galactic RM structure is not the limiting factor for improving estimates of $B_\mathrm{host}$ for FRBs \citep[for recent examples of studies examining the $\mathrm{DM}_\mathrm{host}$, $\mathrm{RM}_\mathrm{host}$, and $\mathrm{B}_\mathrm{host}$ contributed by FRB host galaxies in simulations and observations, see][]{2023MNRAS.518..539M, 2023ApJ...954..179M, 2024A&A...690A..47K, 2025JCAP...01..036A}. Instead, the primary bottlenecks in accurately measuring the FRB host galaxy $B_\mathrm{host}$ are the lack of redshift information for a large fraction of FRBs and the large uncertainties in $\mathrm{DM}_\mathrm{MW}$ and $\mathrm{DM}_\mathrm{IGM}$.

In the case of FRB 20190213B, the estimated $\left| \mathrm{RM}_\mathrm{EG} \right|$ after subtracting the H22 $\mathrm{RM}_\mathrm{MW}$ estimate is $9.5 \pm 13.5$ to $13.9 \pm 13.7~\mathrm{rad}~\mathrm{m}^{-2}$, which is consistent with $\left| \mathrm{RM}_\mathrm{EG} \right| \sim 0~\mathrm{rad}~\mathrm{m}^{-2}$. With our updated $\mathrm{RM}_\mathrm{MW}$ estimate, the $\left| \mathrm{RM}_\mathrm{EG} \right|$ is now $22.6 \pm 11.7$ to $26.9 \pm 11.8~\mathrm{rad}~\mathrm{m}^{-2}$ and is no longer consistent with $\left| \mathrm{RM}_\mathrm{EG} \right| \sim 0~\mathrm{rad}~\mathrm{m}^{-2}$. Prior to this work, FRB 20190213B could have been reasonably interpreted as an FRB originating in a clean, weakly magnetized environment, but we have shown that it instead has $\left| \mathrm{RM}_\mathrm{EG} \right|$ that is similar to what one might expect from some parts of a MW-like host galaxy disk \citep{2019MNRAS.488.4220H}. The case of FRB 20190213B exemplifies how high density RM grids are essential for understanding the magnetoionic environments of low $\left| \mathrm{RM}_\mathrm{obs} \right|$ FRBs.

Theories of FRB origins often invoke young neutron stars \citep[e.g.,][]{2016MNRAS.458L..19C, 2016MNRAS.462..941L, 2017ApJ...841...14M, 2019MNRAS.485.4091M}, which are expected to be enshrouded by complex circumburst environments, such as supernova remnants or pulsar/magnetar wind nebulae. In this regard, the population of FRBs with low local environment RM contributions are of particular interest, as they may be located in exceptionally low density circumburst environments. It is possible that this subset of FRBs is associated with an older population of compact objects where the circumburst environment has expanded and dissipated, or where the objects were imparted with a significant velocity kick at birth and have moved away from their birthplace over time. Further, FRBs in weakly magnetized environments are powerful probes of the intergalactic magnetic field strength \citep{2016Sci...354.1249R, 2016ApJ...824..105A} and can help to answer questions about the origins and evolution of cosmic magnetism.

\subsection{Prospects with upcoming surveys} \label{sec:prospects_discussion}
The blueprint for reconstructing small-scale Galactic RM fluctuations around FRB positions outlined in this work can be applied directly to any FRB with full Stokes data that falls within the sky coverage of a high source density RM survey. Above a declination of $-40~\mathrm{deg}$, VLASS is expected to provide a catalog of radio galaxy RMs with an average grid density of $\sim 7$ polarized sources per square degree \citep{2014arXiv1401.1875M}. While the density of the RM grids from VLASS will not match those presented in this work, they will still probe angular scales that are a factor of $\sim 7$ smaller than is currently possible across the northern sky. Prospects in the southern hemisphere are even more promising, with POSSUM cataloging polarized radio galaxies over $20,000~\mathrm{deg}^{2}$ below a declination of $0~\mathrm{deg}$, with a RM grid density of $\sim 30-40$ sources per square degree \citep{2024AJ....167..226V, Gaensler2024}. Especially for those FRBs within the latter survey's footprint, we will be able to accurately isolate local environment RM contributions to FRBs and potentially identify a population of FRBs from low density and weakly magnetized environments that are ideal cosmological probes of intergalactic magnetic fields.

\section{Conclusions} \label{sec:conclusions}
We have conducted two observing campaigns with the Karl G.~Jansky Very Large Array to observe the polarization properties of background radio galaxies around eight fast radio burst (FRB) sources at high Galactic latitudes. We apply RM-synthesis on each radio galaxy, deriving RMs and quantifying Faraday complexity in each source that exceeds a linearly polarized signal-to-noise of 8. From the linearly polarized radio sources, we construct RM grids with densities of $9-28$ sources per square degree around each of the eight target FRBs. We input these RM grids into a Bayesian inference framework, following the steps described by \cite{affan}, to reconstruct continuous maps of the Galactic RM in regions around each FRB sky position. We compare our reconstructed Galactic RM to the current state of the art Galactic RM sky by \cite{2022A&A...657A..43H}. Finally, we recalculate the RM contributed by the FRB host galaxy and circumburst environment after subtracting our newly computed Galactic RMs towards each FRB. The key conclusions of this work are as follows:
\begin{enumerate}
\item In six of the eight fields, we find previously unresolved Galactic RM amplitude variations on angular scales smaller than $1~\mathrm{deg}^{2}$. In five of these fields, we see sharp Galactic RM fluctuations while in one other field we derive a more granular picture of the smooth large-scale RM gradient across the region. We also potentially identify changes in the sign of the RM in two of these fields. Importantly, all of our fields are at high Galactic latitudes ($|b| \geq 25~\mathrm{deg}$) and we predict that even larger Galactic RM amplitude variations over similar, or smaller, scales are present in regions near the Galactic plane.

\item Towards the FRBs themselves, we see differences of a few$~\mathrm{rad}~\mathrm{m}^{-2}$ up to $\sim 40~\mathrm{rad}~\mathrm{m}^{-2}$ in the Galactic RM estimate between our results and those of \cite{2022A&A...657A..43H}. The statistical uncertainties on our Galactic RM estimates also provide a factor of $\sim 2 - 6$ improvement on those by \cite{2022A&A...657A..43H}. After adjusting the Galactic RM estimate towards FRB 20190213B, we find that its observer frame host galaxy RM is no longer consistent with $0~\mathrm{rad}~\mathrm{m}^{-2}$. We suspect that the host galaxy/local environment RM for as much as $30$\% of all FRB sources (namely those with low observed RM amplitudes) may be incorrectly interpreted by the currently available, low density RM grids over most of the sky. Since variations in the Galactic RM are expected to be larger in magnitude, and over smaller angular scales, near the Galactic plane, it is possible that $30$\% is only a lower limit and an even higher fraction of the polarized FRB population is impacted.
\end{enumerate}

With upcoming radio surveys of polarized radio galaxies in the northern (the Very Large Array Sky Survey) and southern (the Polarization Sky Survey of the Universe's Magnetism) skies, the type of analysis presented in this work will be routinely possible for hundreds to thousands of FRBs, and it will enable us to better understand the progenitors of FRBs from clean magnetoionic environments and identify FRBs that can best probe intergalactic magnetic fields.

\section*{Acknowledgements}
We are grateful to Jamie Farnes, who played a key role in conceiving this project and in obtaining the observations and data.
We thank the anonymous reviewer for careful reading of the manuscript and for their constructive feedback.
We also thank Shannon Vanderwoude, Amanda Cook, and Erik Osinga for useful feedback on the manuscript. 
The University of Toronto operates on the traditional land of the Huron-Wendat, the Seneca, and most recently, the Mississaugas of the Credit River; we are grateful to have the opportunity to work on this land. 
The National Radio Astronomy Observatory and Green Bank Observatory are facilities of the U.S. National
Science Foundation operated under cooperative agreement by Associated Universities, Inc.
A.P. is funded by the NSERC Canada Graduate Scholarshops--Doctoral program.
B.M.G. acknowledges the support of the Natural Sciences and Engineering Research Council of Canada (NSERC) through grant RGPIN-2022-03163, and of the Canada Research Chairs program.
Z.P. is supported by an NWO Veni fellowship (VI.Veni.222.295).
S.P.O. acknowledges support from the Comunidad de Madrid Atracción de Talento program via grant 2022-T1/TIC-23797, and grant PID2023-146372OB-I00 funded by MICIU/AEI/10.13039/501100011033 and by ERDF, EU.
Co-funded by the European Union (ERC, ISM-FLOW, 101055318). Views and opinions expressed are, however, those of the author(s) only and do not necessarily reflect those of the European Union or the European Research Council. Neither the European Union nor the granting authority can be held responsible for them.
Research at Perimeter Institute is supported in part by the Government of Canada through the Department of Innovation, Science and Economic Development and by the Province of Ontario through the Ministry of Colleges and Universities.

\facilities{VLA}

\software{Astropy \citep{2013A&A...558A..33A, 2018AJ....156..123A, 2022ApJ...935..167A},
CASA \citep{2022PASP..134k4501C},
Matplotlib \citep{Hunter:2007}, 
NumPy \citep{harris2020array},
PyGEDM \citep{2021PASA...38...38P},
RM-CLEAN \citep{2009A&A...503..409H},
RM-synthesis \citep{2005A&A...441.1217B}, 
RM-tools \citep{2020ascl.soft05003P},
SciPy \citep{2020SciPy-NMeth}.}

\appendix

\section{Rotation measure catalog of polarized radio sources} \label{sec:appendix}
In Table \ref{tb:catalog}, we tabulate the full RM catalog across all eight fields studied in this work. For each linearly polarized radio source, we provide its sky position in degrees, the $\mathrm{RM}_\mathrm{obs}$ with associated measurement uncertainty $\delta(\mathrm{RM}_\mathrm{obs})$, the Faraday complexity metric $m_2$, and a flag to indicate whether the radio source is Faraday simple (``Y'') or not (``N''). Note that for values of $m_2 < 10^{-2}$, we report $m_2 = 0.00$. The sky positions provided are of the brightest pixel across a given source along which we apply the RM-synthesis analysis and, therefore, does not necessarily correspond to the physical center of the radio galaxy.

\setlength{\tabcolsep}{10pt}
\setlength{\LTcapwidth}{1.0\textwidth}
\begin{center}
\begin{longtable*}{cccccc}
\caption{Rotation measure catalog of all linearly polarized radio sources observed in this work.} \label{tb:catalog}\\
\hline
\hline
J2000 Right Ascension & J2000 Declination & $\mathrm{RM}_\mathrm{obs}$ & $\delta(\mathrm{RM}_\mathrm{obs})$ & $m_2$ & Faraday simple?\\
(deg) & (deg) & (rad~m$^{-2}$) & (rad~m$^{-2}$) &  &  \\
\hline
\endfirsthead

\multicolumn{6}{c}
{{\bfseries \tablename\ \thetable{} -- continued from previous page}} \\
\hline
J2000 Right Ascension & J2000 Declination & $\mathrm{RM}_\mathrm{obs}$ & $\delta(\mathrm{RM}_\mathrm{obs})$ & $m_2$ & Faraday simple?\\
(deg) & (deg) & (rad~m$^{-2}$) & (rad~m$^{-2}$) &  &  \\
\hline
\endhead

\hline \multicolumn{2}{c}{{Continued on next page}} \\ \hline
\endfoot
\hline
\endlastfoot

\hline \multicolumn{6}{c}{Field: FRB 20110523A}\\ \hline
$326.952$ & $-0.491$ & $29.9$ & $7.3$ & $0.00$ & Y\\
$327.030$ & $-0.815$ & $51$ & $12$ & $0.00$ & Y\\
$326.171$ & $0.450$ & $10.4$ & $3.9$ & $0.00$ & Y\\
$326.263$ & $0.535$ & $13.7$ & $4.2$ & $0.00$ & Y\\
$325.970$ & $-0.142$ & $9.3$ & $5.5$ & $0.00$ & Y\\
$325.970$ & $-0.466$ & $-26.2$ & $7.3$ & $0.00$ & Y\\
$325.665$ & $-0.904$ & $21$ & $12$ & $0.00$ & Y\\ 
\hline \multicolumn{6}{c}{Field: FRB 20160102A}\\ \hline
$339.285$ & $-30.346$ & $24.4$ & $2.7$ & $0.00$ & Y\\
$339.286$ & $-30.342$ & $23.4$ & $1.3$ & $0.15$ & N\\
$339.310$ & $-30.088$ & $39$ & $13$ & $0.00$ & Y\\
$339.362$ & $-30.190$ & $18.7$ & $5.3$ & $0.00$ & Y\\
$339.420$ & $-29.849$ & $-2.8$ & $6.4$ & $0.00$ & Y\\
$339.525$ & $-30.100$ & $36.1$ & $8.5$ & $0.00$ & Y\\
$339.589$ & $-30.346$ & $-39$ & $14$ & $0.00$ & Y\\
$339.678$ & $-30.114$ & $41.5$ & $3.0$ & $0.00$ & Y\\
$339.771$ & $-29.707$ & $23.0$ & $4.7$ & $0.00$ & Y\\
$339.794$ & $-30.616$ & $11.9$ & $3.4$ & $0.00$ & Y\\
$339.797$ & $-30.619$ & $23$ & $12$ & $0.00$ & Y\\
$339.894$ & $-30.072$ & $40.6$ & $8.3$ & $0.00$ & Y\\
$339.988$ & $-30.378$ & $18.58$ & $0.86$ & $0.00$ & Y\\
$339.987$ & $-30.380$ & $34.5$ & $1.7$ & $0.00$ & Y\\
$340.078$ & $-30.365$ & $40.7$ & $6.7$ & $0.00$ & Y\\
\hline \multicolumn{6}{c}{Field: FRB 20181030A}\\ \hline
$157.525$ & $73.424$ & $786$ & $12$ & $0.00$ & Y\\
$157.640$ & $73.978$ & $-26$ & $11$ & $0.00$ & Y\\
$157.969$ & $73.426$ & $-28$ & $24$ & $0.10$ & N\\
$158.006$ & $73.429$ & $-21$ & $23$ & $0.10$ & N\\
$158.073$ & $74.063$ & $-55$ & $71$ & $0.23$ & N\\
$159.217$ & $73.655$ & $-13.5$ & $8.7$ & $0.00$ & Y\\
$159.904$ & $73.429$ & $-30$ & $59$ & $0.24$ & N\\
$160.243$ & $73.919$ & $1.9$ & $8.4$ & $0.00$ & Y\\
\hline \multicolumn{6}{c}{Field: FRB 20190117A}\\ \hline
$331.275$ & $17.125$ & $54$ & $10$ & $0.00$ & Y\\
$331.312$ & $17.181$ & $5.6$ & $8.9$ & $0.00$ & Y\\
$331.444$ & $17.210$ & $-39.7$ & $5.4$ & $0.00$ & Y\\
$331.447$ & $17.210$ & $-33$ & $12$ & $0.00$ & Y\\
$331.713$ & $17.172$ & $-37.4$ & $6.3$ & $0.00$ & Y\\
$331.744$ & $17.159$ & $-33.8$ & $2.6$ & $0.00$ & Y\\
$331.782$ & $17.374$ & $-51.4$ & $4.1$ & $0.00$ & Y\\
$331.783$ & $17.374$ & $-54$ & $52$ & $0.22$ & N\\
$331.849$ & $17.596$ & $-34.5$ & $2.7$ & $0.00$ & Y\\
$331.958$ & $17.208$ & $-69.9$ & $9.3$ & $0.00$ & Y\\
$332.051$ & $17.564$ & $-48.3$ & $2.8$ & $0.00$ & Y\\
$332.055$ & $17.564$ & $-20$ & $37$ & $0.16$ & N\\
$332.084$ & $17.549$ & $-34$ & $13$ & $0.00$ & Y\\
$332.163$ & $17.354$ & $-39.8$ & $2.6$ & $0.00$ & Y\\
$331.956$ & $17.204$ & $-78$ & $12$ & $0.00$ & Y\\
$331.706$ & $17.173$ & $-59$ & $13$ & $0.00$ & Y\\
\hline \multicolumn{6}{c}{Field: FRB 20190213B}\\ \hline
$274.102$ & $81.477$ & $12.2$ & $3.8$ & $0.00$ & Y\\
$274.580$ & $81.422$ & $-18.0$ & $7.7$ & $0.00$ & Y\\
$275.879$ & $81.806$ & $44.2$ & $1.1$ & $0.09$ & Y\\
$275.899$ & $81.810$ & $52.2$ & $5.3$ & $0.00$ & Y\\
$275.919$ & $81.415$ & $9.3$ & $5.4$ & $0.00$ & Y\\
$275.956$ & $81.414$ & $22.9$ & $2.5$ & $0.00$ & Y\\
$276.043$ & $81.229$ & $-2.5$ & $8.0$ & $0.00$ & Y\\
$276.255$ & $81.740$ & $36$ & $11$ & $0.00$ & Y\\
$276.412$ & $81.306$ & $-394$ & $54$ & $0.24$ & N\\
$277.431$ & $81.341$ & $3$ & $37$ & $0.16$ & N\\
$277.562$ & $80.987$ & $20.3$ & $7.5$ & $0.00$ & Y\\
$277.576$ & $80.998$ & $0.5$ & $7.3$ & $0.00$ & Y\\
\hline \multicolumn{6}{c}{Field: FRB 20190608B}\\ \hline
$333.589$ & $-8.053$ & $3$ & $10$ & $0.00$ & Y\\
$333.601$ & $-8.083$ & $-46.1$ & $7.8$ & $0.00$ & Y\\
$333.644$ & $-8.084$ & $-36$ & $31$ & $0.13$ & N\\
$333.869$ & $-8.329$ & $-13.27$ & $0.71$ & $0.00$ & Y\\
$333.955$ & $-7.960$ & $0.7$ & $8.0$ & $0.00$ & Y\\
$333.956$ & $-7.961$ & $-30.4$ & $4.0$ & $0.00$ & Y\\
$334.038$ & $-8.276$ & $-24.3$ & $2.9$ & $0.00$ & Y\\
$334.042$ & $-8.278$ & $-31$ & $43$ & $0.19$ & N\\
$334.051$ & $-7.875$ & $-21$ & $29$ & $0.12$ & N\\
$334.081$ & $-8.252$ & $-11.4$ & $8.4$ & $0.00$ & Y\\
\hline \multicolumn{6}{c}{Field: FRB 20191108A}\\ \hline
$22.908$ & $31.793$ & $-82$ & $12$ & $0.00$ & Y\\
$23.011$ & $31.947$ & $-67$ & $11$ & $0.00$ & Y\\
$23.011$ & $31.948$ & $-86$ & $47$ & $0.20$ & N\\
$23.037$ & $31.949$ & $-64.1$ & $7.8$ & $0.00$ & Y \\
$23.322$ & $31.874$ & $-41.7$ & $6.0$ & $0.00$ & Y\\
$23.492$ & $31.721$ & $-33.2$ & $3.1$ & $0.00$ & Y\\
$23.522$ & $32.074$ & $-96$ & $12$ & $0.00$ & Y\\
$23.669$ & $32.079$ & $-31.7$ & $8.6$ & $0.00$ & Y\\
$22.904$ & $31.790$ & $-49$ & $12$ & $0.00$ & Y\\
\hline \multicolumn{6}{c}{Field: FRB 20200120E}\\ \hline
$148.300$ & $69.068$ & $-55$ & $33$ & $0.14$ & N\\
$148.550$ & $68.982$ & $-43.1$ & $5.1$ & $0.00$ & Y\\
$148.888$ & $69.065$ & $-22.9$ & $2.1$ & $0.00$ & Y\\
$149.091$ & $68.766$ & $-4.4$ & $7.2$ & $0.00$ & Y\\
$149.095$ & $68.767$ & $-53.6$ & $4.6$ & $0.00$ & Y\\
$149.426$ & $68.742$ & $2$ & $68$ & $0.24$ & N\\
$149.450$ & $69.175$ & $-32$ & $71$ & $0.21$ & N\\
$149.452$ & $69.180$ & $-13.3$ & $1.5$ & $0.05$ & Y\\
$149.745$ & $68.768$ & $-14.3$ & $5.2$ & $0.00$ & Y\\
$149.749$ & $68.539$ & $-19.8$ & $1.6$ & $0.00$ & Y\\
$149.774$ & $68.548$ & $-31.0$ & $1.7$ & $0.00$ & Y\\
$149.829$ & $68.567$ & $-7.5$ & $5.9$ & $0.00$ & Y\\
$149.832$ & $68.563$ & $50$ & $91$ & $0.23$ & N\\
$150.002$ & $68.523$ & $31.0$ & $6.9$ & $0.20$ & N\\
$150.038$ & $68.537$ & $-8$ & $50$ & $0.22$ & N\\
$150.072$ & $68.652$ & $-1308$ & $14$ & $0.00$ & Y\\
$150.276$ & $68.835$ & $-8.7$ & $2.0$ & $0.07$ & Y\\
$150.285$ & $68.834$ & $-5.0$ & $1.2$ & $0.08$ & Y\\
$150.537$ & $68.978$ & $-628$ & $38$ & $0.19$ & N\\
$150.537$ & $68.977$ & $-330$ & $43$ & $0.18$ & N\\
$150.609$ & $68.973$ & $-53.5$ & $3.0$ & $0.00$ & Y\\
$150.615$ & $68.975$ & $-4.0$ & $1.5$ & $0.00$ & Y\\
\hline
\hline
\end{longtable*}
\end{center}

\bibliography{ref}{}

\begin{thebibliography}{}
\expandafter\ifx\csname natexlab\endcsname\relax\def\natexlab#1{#1}\fi
\providecommand{\url}[1]{\href{#1}{#1}}
\providecommand{\dodoi}[1]{doi:~\href{http://doi.org/#1}{\nolinkurl{#1}}}
\providecommand{\doeprint}[1]{\href{http://ascl.net/#1}{\nolinkurl{http://ascl.net/#1}}}
\providecommand{\doarXiv}[1]{\href{https://arxiv.org/abs/#1}{\nolinkurl{https://arxiv.org/abs/#1}}}

\bibitem[{{Acharya} \& {Beniamini}(2025)}]{2025JCAP...01..036A}
{Acharya}, S.~K., \& {Beniamini}, P. 2025, \jcap, 2025, 036, \dodoi{10.1088/1475-7516/2025/01/036}

\bibitem[{{Akahori} {et~al.}(2016){Akahori}, {Ryu}, \& {Gaensler}}]{2016ApJ...824..105A}
{Akahori}, T., {Ryu}, D., \& {Gaensler}, B.~M. 2016, \apj, 824, 105, \dodoi{10.3847/0004-637X/824/2/105}

\bibitem[{{Astropy Collaboration} {et~al.}(2013){Astropy Collaboration}, {Robitaille}, {Tollerud}, {Greenfield}, {Droettboom}, {Bray}, {Aldcroft}, {Davis}, {Ginsburg}, {Price-Whelan}, {Kerzendorf}, {Conley}, {Crighton}, {Barbary}, {Muna}, {Ferguson}, {Grollier}, {Parikh}, {Nair}, {Unther}, {Deil}, {Woillez}, {Conseil}, {Kramer}, {Turner}, {Singer}, {Fox}, {Weaver}, {Zabalza}, {Edwards}, {Azalee Bostroem}, {Burke}, {Casey}, {Crawford}, {Dencheva}, {Ely}, {Jenness}, {Labrie}, {Lim}, {Pierfederici}, {Pontzen}, {Ptak}, {Refsdal}, {Servillat}, \& {Streicher}}]{2013A&A...558A..33A}
{Astropy Collaboration}, {Robitaille}, T.~P., {Tollerud}, E.~J., {et~al.} 2013, \aap, 558, A33, \dodoi{10.1051/0004-6361/201322068}

\bibitem[{{Astropy Collaboration} {et~al.}(2018){Astropy Collaboration}, {Price-Whelan}, {Sip{\H{o}}cz}, {G{\"u}nther}, {Lim}, {Crawford}, {Conseil}, {Shupe}, {Craig}, {Dencheva}, {Ginsburg}, {VanderPlas}, {Bradley}, {P{\'e}rez-Su{\'a}rez}, {de Val-Borro}, {Aldcroft}, {Cruz}, {Robitaille}, {Tollerud}, {Ardelean}, {Babej}, {Bach}, {Bachetti}, {Bakanov}, {Bamford}, {Barentsen}, {Barmby}, {Baumbach}, {Berry}, {Biscani}, {Boquien}, {Bostroem}, {Bouma}, {Brammer}, {Bray}, {Breytenbach}, {Buddelmeijer}, {Burke}, {Calderone}, {Cano Rodr{\'\i}guez}, {Cara}, {Cardoso}, {Cheedella}, {Copin}, {Corrales}, {Crichton}, {D'Avella}, {Deil}, {Depagne}, {Dietrich}, {Donath}, {Droettboom}, {Earl}, {Erben}, {Fabbro}, {Ferreira}, {Finethy}, {Fox}, {Garrison}, {Gibbons}, {Goldstein}, {Gommers}, {Greco}, {Greenfield}, {Groener}, {Grollier}, {Hagen}, {Hirst}, {Homeier}, {Horton}, {Hosseinzadeh}, {Hu}, {Hunkeler}, {Ivezi{\'c}}, {Jain}, {Jenness}, {Kanarek}, {Kendrew}, {Kern}, {Kerzendorf}, {Khvalko}, {King}, {Kirkby}, {Kulkarni},
  {Kumar}, {Lee}, {Lenz}, {Littlefair}, {Ma}, {Macleod}, {Mastropietro}, {McCully}, {Montagnac}, {Morris}, {Mueller}, {Mumford}, {Muna}, {Murphy}, {Nelson}, {Nguyen}, {Ninan}, {N{\"o}the}, {Ogaz}, {Oh}, {Parejko}, {Parley}, {Pascual}, {Patil}, {Patil}, {Plunkett}, {Prochaska}, {Rastogi}, {Reddy Janga}, {Sabater}, {Sakurikar}, {Seifert}, {Sherbert}, {Sherwood-Taylor}, {Shih}, {Sick}, {Silbiger}, {Singanamalla}, {Singer}, {Sladen}, {Sooley}, {Sornarajah}, {Streicher}, {Teuben}, {Thomas}, {Tremblay}, {Turner}, {Terr{\'o}n}, {van Kerkwijk}, {de la Vega}, {Watkins}, {Weaver}, {Whitmore}, {Woillez}, {Zabalza}, \& {Astropy Contributors}}]{2018AJ....156..123A}
{Astropy Collaboration}, {Price-Whelan}, A.~M., {Sip{\H{o}}cz}, B.~M., {et~al.} 2018, \aj, 156, 123, \dodoi{10.3847/1538-3881/aabc4f}

\bibitem[{{Astropy Collaboration} {et~al.}(2022){Astropy Collaboration}, {Price-Whelan}, {Lim}, {Earl}, {Starkman}, {Bradley}, {Shupe}, {Patil}, {Corrales}, {Brasseur}, {N{\"o}the}, {Donath}, {Tollerud}, {Morris}, {Ginsburg}, {Vaher}, {Weaver}, {Tocknell}, {Jamieson}, {van Kerkwijk}, {Robitaille}, {Merry}, {Bachetti}, {G{\"u}nther}, {Aldcroft}, {Alvarado-Montes}, {Archibald}, {B{\'o}di}, {Bapat}, {Barentsen}, {Baz{\'a}n}, {Biswas}, {Boquien}, {Burke}, {Cara}, {Cara}, {Conroy}, {Conseil}, {Craig}, {Cross}, {Cruz}, {D'Eugenio}, {Dencheva}, {Devillepoix}, {Dietrich}, {Eigenbrot}, {Erben}, {Ferreira}, {Foreman-Mackey}, {Fox}, {Freij}, {Garg}, {Geda}, {Glattly}, {Gondhalekar}, {Gordon}, {Grant}, {Greenfield}, {Groener}, {Guest}, {Gurovich}, {Handberg}, {Hart}, {Hatfield-Dodds}, {Homeier}, {Hosseinzadeh}, {Jenness}, {Jones}, {Joseph}, {Kalmbach}, {Karamehmetoglu}, {Ka{\l}uszy{\'n}ski}, {Kelley}, {Kern}, {Kerzendorf}, {Koch}, {Kulumani}, {Lee}, {Ly}, {Ma}, {MacBride}, {Maljaars}, {Muna}, {Murphy}, {Norman},
  {O'Steen}, {Oman}, {Pacifici}, {Pascual}, {Pascual-Granado}, {Patil}, {Perren}, {Pickering}, {Rastogi}, {Roulston}, {Ryan}, {Rykoff}, {Sabater}, {Sakurikar}, {Salgado}, {Sanghi}, {Saunders}, {Savchenko}, {Schwardt}, {Seifert-Eckert}, {Shih}, {Jain}, {Shukla}, {Sick}, {Simpson}, {Singanamalla}, {Singer}, {Singhal}, {Sinha}, {Sip{\H{o}}cz}, {Spitler}, {Stansby}, {Streicher}, {{\v{S}}umak}, {Swinbank}, {Taranu}, {Tewary}, {Tremblay}, {de Val-Borro}, {Van Kooten}, {Vasovi{\'c}}, {Verma}, {de Miranda Cardoso}, {Williams}, {Wilson}, {Winkel}, {Wood-Vasey}, {Xue}, {Yoachim}, {Zhang}, {Zonca}, \& {Astropy Project Contributors}}]{2022ApJ...935..167A}
{Astropy Collaboration}, {Price-Whelan}, A.~M., {Lim}, P.~L., {et~al.} 2022, \apj, 935, 167, \dodoi{10.3847/1538-4357/ac7c74}

\bibitem[{{Bannister} {et~al.}(2019){Bannister}, {Deller}, {Phillips}, {Macquart}, {Prochaska}, {Tejos}, {Ryder}, {Sadler}, {Shannon}, {Simha}, {Day}, {McQuinn}, {North-Hickey}, {Bhandari}, {Arcus}, {Bennert}, {Burchett}, {Bouwhuis}, {Dodson}, {Ekers}, {Farah}, {Flynn}, {James}, {Kerr}, {Lenc}, {Mahony}, {O'Meara}, {Os{\l}owski}, {Qiu}, {Treu}, {U}, {Bateman}, {Bock}, {Bolton}, {Brown}, {Bunton}, {Chippendale}, {Cooray}, {Cornwell}, {Gupta}, {Hayman}, {Kesteven}, {Koribalski}, {MacLeod}, {McClure-Griffiths}, {Neuhold}, {Norris}, {Pilawa}, {Qiao}, {Reynolds}, {Roxby}, {Shimwell}, {Voronkov}, \& {Wilson}}]{2019Sci...365..565B}
{Bannister}, K.~W., {Deller}, A.~T., {Phillips}, C., {et~al.} 2019, Science, 365, 565, \dodoi{10.1126/science.aaw5903}

\bibitem[{{Baptista} {et~al.}(2024){Baptista}, {Prochaska}, {Mannings}, {James}, {Shannon}, {Ryder}, {Deller}, {Scott}, {Glowacki}, \& {Tejos}}]{2024ApJ...965...57B}
{Baptista}, J., {Prochaska}, J.~X., {Mannings}, A.~G., {et~al.} 2024, \apj, 965, 57, \dodoi{10.3847/1538-4357/ad2705}

\bibitem[{{Beck}(2001)}]{2001SSRv...99..243B}
{Beck}, R. 2001, \ssr, 99, 243, \dodoi{10.1023/A:1013805401252}

\bibitem[{{Bhandari} {et~al.}(2018){Bhandari}, {Keane}, {Barr}, {Jameson}, {Petroff}, {Johnston}, {Bailes}, {Bhat}, {Burgay}, {Burke-Spolaor}, {Caleb}, {Eatough}, {Flynn}, {Green}, {Jankowski}, {Kramer}, {Krishnan}, {Morello}, {Possenti}, {Stappers}, {Tiburzi}, {van Straten}, {Andreoni}, {Butterley}, {Chandra}, {Cooke}, {Corongiu}, {Coward}, {Dhillon}, {Dodson}, {Hardy}, {Howell}, {Jaroenjittichai}, {Klotz}, {Littlefair}, {Marsh}, {Mickaliger}, {Muxlow}, {Perrodin}, {Pritchard}, {Sawangwit}, {Terai}, {Tominaga}, {Torne}, {Totani}, {Trois}, {Turpin}, {Niino}, {Wilson}, {Albert}, {Andr{\'e}}, {Anghinolfi}, {Anton}, {Ardid}, {Aubert}, {Avgitas}, {Baret}, {Barrios-Mart{\'\i}}, {Basa}, {Belhorma}, {Bertin}, {Biagi}, {Bormuth}, {Bourret}, {Bouwhuis}, {Br{\^a}nza{\c{s}}}, {Bruijn}, {Brunner}, {Busto}, {Capone}, {Caramete}, {Carr}, {Celli}, {Moursli}, {Chiarusi}, {Circella}, {Coelho}, {Coleiro}, {Coniglione}, {Costantini}, {Coyle}, {Creusot}, {D{\'\i}az}, {Deschamps}, {De Bonis}, {Distefano}, {Palma}, {Domi},
  {Donzaud}, {Dornic}, {Drouhin}, {Eberl}, {Bojaddaini}, {Khayati}, {Els{\"a}sser}, {Enzenh{\"o}fer}, {Ettahiri}, {Fassi}, {Felis}, {Fusco}, {Gay}, {Giordano}, {Glotin}, {Gregoire}, {Gracia-Ruiz}, {Graf}, {Hallmann}, {van Haren}, {Heijboer}, {Hello}, {Hern{\'a}ndez-Rey}, {H{\"o}{\ss}l}, {Hofest{\"a}dt}, {Hugon}, {Illuminati}, {James}, {de Jong}, {Jongen}, {Kadler}, {Kalekin}, {Katz}, {Kie{\ss}ling}, {Kouchner}, {Kreter}, {Kreykenbohm}, {Kulikovskiy}, {Lachaud}, {Lahmann}, {Lef{\`e}vre}, {Leonora}, {Loucatos}, {Marcelin}, {Margiotta}, {Marinelli}, {Mart{\'\i}nez-Mora}, {Mele}, {Melis}, {Michael}, {Migliozzi}, {Moussa}, {Navas}, {Nezri}, {Organokov}, {P{\v{a}}v{\v{a}}la{\c{s}}}, {Pellegrino}, {Perrina}, {Piattelli}, {Popa}, {Pradier}, {Quinn}, {Racca}, {Riccobene}, {S{\'a}nchez-Losa}, {Salda{\~n}a}, {Salvadori}, {Samtleben}, {Sanguineti}, {Sapienza}, {Sch{\"u}ssler}, {Sieger}, {Spurio}, {Stolarczyk}, {Taiuti}, {Tayalati}, {Trovato}, {Turpin}, {T{\"o}nnis}, {Vallage}, {Van Elewyck}, {Versari}, {Vivolo},
  {Vizzocca}, {Wilms}, {Zornoza}, \& {Z{\'u}{\~n}iga}}]{2018MNRAS.475.1427B}
{Bhandari}, S., {Keane}, E.~F., {Barr}, E.~D., {et~al.} 2018, \mnras, 475, 1427, \dodoi{10.1093/mnras/stx3074}

\bibitem[{{Bhandari} {et~al.}(2020{\natexlab{a}}){Bhandari}, {Sadler}, {Prochaska}, {Simha}, {Ryder}, {Marnoch}, {Bannister}, {Macquart}, {Flynn}, {Shannon}, {Tejos}, {Corro-Guerra}, {Day}, {Deller}, {Ekers}, {Lopez}, {Mahony}, {Nu{\~n}ez}, \& {Phillips}}]{2020ApJ...895L..37B}
{Bhandari}, S., {Sadler}, E.~M., {Prochaska}, J.~X., {et~al.} 2020{\natexlab{a}}, \apjl, 895, L37, \dodoi{10.3847/2041-8213/ab672e}

\bibitem[{{Bhandari} {et~al.}(2020{\natexlab{b}}){Bhandari}, {Bannister}, {Lenc}, {Cho}, {Ekers}, {Day}, {Deller}, {Flynn}, {James}, {Macquart}, {Mahony}, {Marnoch}, {Moss}, {Phillips}, {Prochaska}, {Qiu}, {Ryder}, {Shannon}, {Tejos}, \& {Wong}}]{2020ApJ...901L..20B}
{Bhandari}, S., {Bannister}, K.~W., {Lenc}, E., {et~al.} 2020{\natexlab{b}}, \apjl, 901, L20, \dodoi{10.3847/2041-8213/abb462}

\bibitem[{{Bhardwaj} {et~al.}(2021{\natexlab{a}}){Bhardwaj}, {Kirichenko}, {Michilli}, {Mayya}, {Kaspi}, {Gaensler}, {Rahman}, {Tendulkar}, {Fonseca}, {Josephy}, {Leung}, {Merryfield}, {Petroff}, {Pleunis}, {Sanghavi}, {Scholz}, {Shin}, {Smith}, \& {Stairs}}]{2021ApJ...919L..24B}
{Bhardwaj}, M., {Kirichenko}, A.~Y., {Michilli}, D., {et~al.} 2021{\natexlab{a}}, \apjl, 919, L24, \dodoi{10.3847/2041-8213/ac223b}

\bibitem[{{Bhardwaj} {et~al.}(2021{\natexlab{b}}){Bhardwaj}, {Gaensler}, {Kaspi}, {Landecker}, {Mckinven}, {Michilli}, {Pleunis}, {Tendulkar}, {Andersen}, {Boyle}, {Cassanelli}, {Chawla}, {Cook}, {Dobbs}, {Fonseca}, {Kaczmarek}, {Leung}, {Masui}, {Mnchmeyer}, {Ng}, {Rafiei-Ravandi}, {Scholz}, {Shin}, {Smith}, {Stairs}, \& {Zwaniga}}]{2021ApJ...910L..18B}
{Bhardwaj}, M., {Gaensler}, B.~M., {Kaspi}, V.~M., {et~al.} 2021{\natexlab{b}}, \apjl, 910, L18, \dodoi{10.3847/2041-8213/abeaa6}

\bibitem[{{Bhardwaj} {et~al.}(2024){Bhardwaj}, {Michilli}, {Kirichenko}, {Modilim}, {Shin}, {Kaspi}, {Andersen}, {Cassanelli}, {Brar}, {Chatterjee}, {Cook}, {Dong}, {Fonseca}, {Gaensler}, {Ibik}, {Kaczmarek}, {Lanman}, {Leung}, {Masui}, {Pandhi}, {Pearlman}, {Petroff}, {Pleunis}, {Prochaska}, {Rafiei-Ravandi}, {Sand}, {Scholz}, \& {Smith}}]{2024ApJ...971L..51B}
{Bhardwaj}, M., {Michilli}, D., {Kirichenko}, A.~Y., {et~al.} 2024, \apjl, 971, L51, \dodoi{10.3847/2041-8213/ad64d1}

\bibitem[{{Brentjens} \& {de Bruyn}(2005)}]{2005A&A...441.1217B}
{Brentjens}, M.~A., \& {de Bruyn}, A.~G. 2005, \aap, 441, 1217, \dodoi{10.1051/0004-6361:20052990}

\bibitem[{{Briggs}(1995)}]{1995PhDT.......238B}
{Briggs}, D.~S. 1995, PhD thesis, New Mexico Institute of Mining and Technology

\bibitem[{{Brown}(2011)}]{brown2011_m2}
{Brown}, S. 2011, {Internal POSSUM Report \#9: Assess the Complexity of an RM Synthesis Spectrum.}
\newblock \url{https://askap.org/possum/wiki/uploads/Documents/possumreport_009.pdf}

\bibitem[{{Burn}(1966)}]{1966MNRAS.133...67B}
{Burn}, B.~J. 1966, \mnras, 133, 67, \dodoi{10.1093/mnras/133.1.67}

\bibitem[{{Caleb} {et~al.}(2018){Caleb}, {Keane}, {van Straten}, {Kramer}, {Macquart}, {Bailes}, {Barr}, {Bhat}, {Bhandari}, {Burgay}, {Farah}, {Jameson}, {Jankowski}, {Johnston}, {Petroff}, {Possenti}, {Stappers}, {Tiburzi}, \& {Venkatraman Krishnan}}]{2018MNRAS.478.2046C}
{Caleb}, M., {Keane}, E.~F., {van Straten}, W., {et~al.} 2018, \mnras, 478, 2046, \dodoi{10.1093/mnras/sty1137}

\bibitem[{{CASA Team} {et~al.}(2022){CASA Team}, {Bean}, {Bhatnagar}, {Castro}, {Donovan Meyer}, {Emonts}, {Garcia}, {Garwood}, {Golap}, {Gonzalez Villalba}, {Harris}, {Hayashi}, {Hoskins}, {Hsieh}, {Jagannathan}, {Kawasaki}, {Keimpema}, {Kettenis}, {Lopez}, {Marvil}, {Masters}, {McNichols}, {Mehringer}, {Miel}, {Moellenbrock}, {Montesino}, {Nakazato}, {Ott}, {Petry}, {Pokorny}, {Raba}, {Rau}, {Schiebel}, {Schweighart}, {Sekhar}, {Shimada}, {Small}, {Steeb}, {Sugimoto}, {Suoranta}, {Tsutsumi}, {van Bemmel}, {Verkouter}, {Wells}, {Xiong}, {Szomoru}, {Griffith}, {Glendenning}, \& {Kern}}]{2022PASP..134k4501C}
{CASA Team}, {Bean}, B., {Bhatnagar}, S., {et~al.} 2022, \pasp, 134, 114501, \dodoi{10.1088/1538-3873/ac9642}

\bibitem[{{Chatterjee} {et~al.}(2017){Chatterjee}, {Law}, {Wharton}, {Burke-Spolaor}, {Hessels}, {Bower}, {Cordes}, {Tendulkar}, {Bassa}, {Demorest}, {Butler}, {Seymour}, {Scholz}, {Abruzzo}, {Bogdanov}, {Kaspi}, {Keimpema}, {Lazio}, {Marcote}, {McLaughlin}, {Paragi}, {Ransom}, {Rupen}, {Spitler}, \& {van Langevelde}}]{2017Natur.541...58C}
{Chatterjee}, S., {Law}, C.~J., {Wharton}, R.~S., {et~al.} 2017, \nat, 541, 58, \dodoi{10.1038/nature20797}

\bibitem[{{CHIME/FRB Collaboration} {et~al.}(2021){CHIME/FRB Collaboration}, {Amiri}, {Andersen}, {Bandura}, {Berger}, {Bhardwaj}, {Boyce}, {Boyle}, {Brar}, {Breitman}, {Cassanelli}, {Chawla}, {Chen}, {Cliche}, {Cook}, {Cubranic}, {Curtin}, {Deng}, {Dobbs}, {Dong}, {Eadie}, {Fandino}, {Fonseca}, {Gaensler}, {Giri}, {Good}, {Halpern}, {Hill}, {Hinshaw}, {Josephy}, {Kaczmarek}, {Kader}, {Kania}, {Kaspi}, {Landecker}, {Lang}, {Leung}, {Li}, {Lin}, {Masui}, {McKinven}, {Mena-Parra}, {Merryfield}, {Meyers}, {Michilli}, {Milutinovic}, {Mirhosseini}, {M{\"u}nchmeyer}, {Naidu}, {Newburgh}, {Ng}, {Patel}, {Pen}, {Petroff}, {Pinsonneault-Marotte}, {Pleunis}, {Rafiei-Ravandi}, {Rahman}, {Ransom}, {Renard}, {Sanghavi}, {Scholz}, {Shaw}, {Shin}, {Siegel}, {Sikora}, {Singh}, {Smith}, {Stairs}, {Tan}, {Tendulkar}, {Vanderlinde}, {Wang}, {Wulf}, \& {Zwaniga}}]{2021ApJS..257...59C}
{CHIME/FRB Collaboration}, {Amiri}, M., {Andersen}, B.~C., {et~al.} 2021, \apjs, 257, 59, \dodoi{10.3847/1538-4365/ac33ab}

\bibitem[{{CHIME/FRB Collaboration} {et~al.}(2024){CHIME/FRB Collaboration}, {Amiri}, {Andersen}, {Andrew}, {Bandura}, {Bhardwaj}, {Boyle}, {Brar}, {Breitman}, {Cassanelli}, {Chawla}, {Cook}, {Curtin}, {Dobbs}, {Dong}, {Eadie}, {Fonseca}, {Gaensler}, {Giri}, {Herrera-Martin}, {Hopkins}, {Ibik}, {Joseph}, {Kaczmarek}, {Kader}, {Kaspi}, {Lanman}, {Lazda}, {Leung}, {Liu}, {Masui}, {McKinven}, {Mena-Parra}, {Merryfield}, {Michilli}, {Ng}, {Nimmo}, {Noble}, {Pandhi}, {Patel}, {Pearlman}, {Pen}, {Petroff}, {Pleunis}, {Rafiei-Ravandi}, {Rahman}, {Ransom}, {Sand}, {Scholz}, {Shah}, {Shin}, {Shpunarska}, {Siegel}, {Smith}, {Stairs}, {Stenning}, {Vanderlinde}, {Wang}, {White}, \& {Wulf}}]{Michilli2024}
---. 2024, \apj, 969, 145, \dodoi{10.3847/1538-4357/ad464b}

\bibitem[{{Chittidi} {et~al.}(2021){Chittidi}, {Simha}, {Mannings}, {Prochaska}, {Ryder}, {Rafelski}, {Neeleman}, {Macquart}, {Tejos}, {Jorgenson}, {Day}, {Marnoch}, {Bhandari}, {Deller}, {Qiu}, {Bannister}, {Shannon}, \& {Heintz}}]{2021ApJ...922..173C}
{Chittidi}, J.~S., {Simha}, S., {Mannings}, A., {et~al.} 2021, \apj, 922, 173, \dodoi{10.3847/1538-4357/ac2818}

\bibitem[{{Condon} {et~al.}(1998){Condon}, {Cotton}, {Greisen}, {Yin}, {Perley}, {Taylor}, \& {Broderick}}]{1998AJ....115.1693C}
{Condon}, J.~J., {Cotton}, W.~D., {Greisen}, E.~W., {et~al.} 1998, \aj, 115, 1693, \dodoi{10.1086/300337}

\bibitem[{{Connor} {et~al.}(2016){Connor}, {Sievers}, \& {Pen}}]{2016MNRAS.458L..19C}
{Connor}, L., {Sievers}, J., \& {Pen}, U.-L. 2016, \mnras, 458, L19, \dodoi{10.1093/mnrasl/slv124}

\bibitem[{{Connor} {et~al.}(2020){Connor}, {van Leeuwen}, {Oostrum}, {Petroff}, {Maan}, {Adams}, {Attema}, {Bast}, {Boersma}, {D{\'e}nes}, {Gardenier}, {Hargreaves}, {Kooistra}, {Pastor-Marazuela}, {Schulz}, {Sclocco}, {Smits}, {Straal}, {van der Schuur}, {Vohl}, {Adebahr}, {de Blok}, {van Cappellen}, {Coolen}, {Damstra}, {van Diepen}, {Frank}, {Hess}, {Hut}, {Kutkin}, {Loose}, {Lucero}, {Mika}, {Moss}, {Mulder}, {Oosterloo}, {Ruiter}, {Vedantham}, {Vermaas}, {Wijnholds}, \& {Ziemke}}]{2020MNRAS.499.4716C}
{Connor}, L., {van Leeuwen}, J., {Oostrum}, L.~C., {et~al.} 2020, \mnras, 499, 4716, \dodoi{10.1093/mnras/staa3009}

\bibitem[{{Cook} {et~al.}(2023){Cook}, {Bhardwaj}, {Gaensler}, {Scholz}, {Eadie}, {Hill}, {Kaspi}, {Masui}, {Curtin}, {Dong}, {Fonseca}, {Herrera-Martin}, {Kaczmarek}, {Lanman}, {Lazda}, {Leung}, {Meyers}, {Michilli}, {Pandhi}, {Pearlman}, {Pleunis}, {Ransom}, {Rahman}, {Sand}, {Shin}, {Smith}, {Stairs}, \& {Stenning}}]{2023ApJ...946...58C}
{Cook}, A.~M., {Bhardwaj}, M., {Gaensler}, B.~M., {et~al.} 2023, \apj, 946, 58, \dodoi{10.3847/1538-4357/acbbd0}

\bibitem[{{Cordes} \& {Lazio}(2002)}]{2002astro.ph..7156C}
{Cordes}, J.~M., \& {Lazio}, T.~J.~W. 2002, arXiv e-prints, astro, \dodoi{10.48550/arXiv.astro-ph/0207156}

\bibitem[{{Day} {et~al.}(2020){Day}, {Deller}, {Shannon}, {Qiu(邱昊)}, {Bannister}, {Bhandari}, {Ekers}, {Flynn}, {James}, {Macquart}, {Mahony}, {Phillips}, \& {Xavier Prochaska}}]{2020MNRAS.497.3335D}
{Day}, C.~K., {Deller}, A.~T., {Shannon}, R.~M., {et~al.} 2020, \mnras, 497, 3335, \dodoi{10.1093/mnras/staa2138}

\bibitem[{{Dolag} {et~al.}(2015){Dolag}, {Gaensler}, {Beck}, \& {Beck}}]{2015MNRAS.451.4277D}
{Dolag}, K., {Gaensler}, B.~M., {Beck}, A.~M., \& {Beck}, M.~C. 2015, \mnras, 451, 4277, \dodoi{10.1093/mnras/stv1190}

\bibitem[{Edenhofer {et~al.}(2024)Edenhofer, Frank, Roth, Leike, Guerdi, Scheel-Platz, Guardiani, Eberle, Westerkamp, \& Enßlin}]{niftyre}
Edenhofer, G., Frank, P., Roth, J., {et~al.} 2024, Journal of Open Source Software, 9, 6593, \dodoi{10.21105/joss.06593}

\bibitem[{{Gaensler} {et~al.}(2024)}]{Gaensler2024}
{Gaensler}, B., {et~al.} 2024, \pasa, submitted

\bibitem[{{Gordon} {et~al.}(2021){Gordon}, {Boyce}, {O'Dea}, {Rudnick}, {Andernach}, {Vantyghem}, {Baum}, {Bui}, {Dionyssiou}, {Safi-Harb}, \& {Sander}}]{2021ApJS..255...30G}
{Gordon}, Y.~A., {Boyce}, M.~M., {O'Dea}, C.~P., {et~al.} 2021, \apjs, 255, 30, \dodoi{10.3847/1538-4365/ac05c0}

\bibitem[{{Hackstein} {et~al.}(2019){Hackstein}, {Br{\"u}ggen}, {Vazza}, {Gaensler}, \& {Heesen}}]{2019MNRAS.488.4220H}
{Hackstein}, S., {Br{\"u}ggen}, M., {Vazza}, F., {Gaensler}, B.~M., \& {Heesen}, V. 2019, \mnras, 488, 4220, \dodoi{10.1093/mnras/stz2033}

\bibitem[{Harris {et~al.}(2020)Harris, Millman, van~der Walt, Gommers, Virtanen, Cournapeau, Wieser, Taylor, Berg, Smith, Kern, Picus, Hoyer, van Kerkwijk, Brett, Haldane, del R{\'{i}}o, Wiebe, Peterson, G{\'{e}}rard-Marchant, Sheppard, Reddy, Weckesser, Abbasi, Gohlke, \& Oliphant}]{harris2020array}
Harris, C.~R., Millman, K.~J., van~der Walt, S.~J., {et~al.} 2020, Nature, 585, 357, \dodoi{10.1038/s41586-020-2649-2}

\bibitem[{{Heald} {et~al.}(2009){Heald}, {Braun}, \& {Edmonds}}]{2009A&A...503..409H}
{Heald}, G., {Braun}, R., \& {Edmonds}, R. 2009, \aap, 503, 409, \dodoi{10.1051/0004-6361/200912240}

\bibitem[{{H{\"o}gbom}(1974)}]{1974A&AS...15..417H}
{H{\"o}gbom}, J.~A. 1974, \aaps, 15, 417

\bibitem[{Hunter(2007)}]{Hunter:2007}
Hunter, J.~D. 2007, Computing in Science \& Engineering, 9, 90, \dodoi{10.1109/MCSE.2007.55}

\bibitem[{{Hutschenreuter} \& {En{\ss}lin}(2020)}]{2020A&A...633A.150H}
{Hutschenreuter}, S., \& {En{\ss}lin}, T.~A. 2020, \aap, 633, A150, \dodoi{10.1051/0004-6361/201935479}

\bibitem[{{Hutschenreuter} {et~al.}(2024){Hutschenreuter}, {Haverkorn}, {Frank}, {Raycheva}, \& {En{\ss}lin}}]{2024A&A...690A.314H}
{Hutschenreuter}, S., {Haverkorn}, M., {Frank}, P., {Raycheva}, N.~C., \& {En{\ss}lin}, T.~A. 2024, \aap, 690, A314, \dodoi{10.1051/0004-6361/202346740}

\bibitem[{{Hutschenreuter} {et~al.}(2022){Hutschenreuter}, {Anderson}, {Betti}, {Bower}, {Brown}, {Br{\"u}ggen}, {Carretti}, {Clarke}, {Clegg}, {Costa}, {Croft}, {Van Eck}, {Gaensler}, {de Gasperin}, {Haverkorn}, {Heald}, {Hull}, {Inoue}, {Johnston-Hollitt}, {Kaczmarek}, {Law}, {Ma}, {MacMahon}, {Mao}, {Riseley}, {Roy}, {Shanahan}, {Shimwell}, {Stil}, {Sobey}, {O'Sullivan}, {Tasse}, {Vacca}, {Vernstrom}, {Williams}, {Wright}, \& {En{\ss}lin}}]{2022A&A...657A..43H}
{Hutschenreuter}, S., {Anderson}, C.~S., {Betti}, S., {et~al.} 2022, \aap, 657, A43, \dodoi{10.1051/0004-6361/202140486}

\bibitem[{{Khadir} {et~al.}(2024){Khadir}, {Pandhi}, {Hutschenreuter}, {Gaensler}, {Vanderwoude}, {West}, \& {O'Sullivan}}]{affan}
{Khadir}, A., {Pandhi}, A., {Hutschenreuter}, S., {et~al.} 2024, arXiv e-prints, arXiv:2410.15265, \dodoi{10.48550/arXiv.2410.15265}

\bibitem[{{Kirsten} {et~al.}(2022){Kirsten}, {Marcote}, {Nimmo}, {Hessels}, {Bhardwaj}, {Tendulkar}, {Keimpema}, {Yang}, {Snelders}, {Scholz}, {Pearlman}, {Law}, {Peters}, {Giroletti}, {Paragi}, {Bassa}, {Hewitt}, {Bach}, {Bezrukovs}, {Burgay}, {Buttaccio}, {Conway}, {Corongiu}, {Feiler}, {Forss{\'e}n}, {Gawro{\'n}ski}, {Karuppusamy}, {Kharinov}, {Lindqvist}, {Maccaferri}, {Melnikov}, {Ould-Boukattine}, {Possenti}, {Surcis}, {Wang}, {Yuan}, {Aggarwal}, {Anna-Thomas}, {Bower}, {Blaauw}, {Burke-Spolaor}, {Cassanelli}, {Clarke}, {Fonseca}, {Gaensler}, {Gopinath}, {Kaspi}, {Kassim}, {Lazio}, {Leung}, {Li}, {Lin}, {Masui}, {Mckinven}, {Michilli}, {Mikhailov}, {Ng}, {Orbidans}, {Pen}, {Petroff}, {Rahman}, {Ransom}, {Shin}, {Smith}, {Stairs}, \& {Vlemmings}}]{2022Natur.602..585K}
{Kirsten}, F., {Marcote}, B., {Nimmo}, K., {et~al.} 2022, \nat, 602, 585, \dodoi{10.1038/s41586-021-04354-w}

\bibitem[{{Kovacs} {et~al.}(2024){Kovacs}, {Mao}, {Basu}, {Ma}, {Pakmor}, {Spitler}, \& {Walker}}]{2024A&A...690A..47K}
{Kovacs}, T.~O., {Mao}, S.~A., {Basu}, A., {et~al.} 2024, \aap, 690, A47, \dodoi{10.1051/0004-6361/202347459}

\bibitem[{{Lacy} {et~al.}(2020){Lacy}, {Baum}, {Chandler}, {Chatterjee}, {Clarke}, {Deustua}, {English}, {Farnes}, {Gaensler}, {Gugliucci}, {Hallinan}, {Kent}, {Kimball}, {Law}, {Lazio}, {Marvil}, {Mao}, {Medlin}, {Mooley}, {Murphy}, {Myers}, {Osten}, {Richards}, {Rosolowsky}, {Rudnick}, {Schinzel}, {Sivakoff}, {Sjouwerman}, {Taylor}, {White}, {Wrobel}, {Andernach}, {Beasley}, {Berger}, {Bhatnager}, {Birkinshaw}, {Bower}, {Brandt}, {Brown}, {Burke-Spolaor}, {Butler}, {Comerford}, {Demorest}, {Fu}, {Giacintucci}, {Golap}, {G{\"u}th}, {Hales}, {Hiriart}, {Hodge}, {Horesh}, {Ivezi{\'c}}, {Jarvis}, {Kamble}, {Kassim}, {Liu}, {Loinard}, {Lyons}, {Masters}, {Mezcua}, {Moellenbrock}, {Mroczkowski}, {Nyland}, {O'Dea}, {O'Sullivan}, {Peters}, {Radford}, {Rao}, {Robnett}, {Salcido}, {Shen}, {Sobotka}, {Witz}, {Vaccari}, {van Weeren}, {Vargas}, {Williams}, \& {Yoon}}]{2020PASP..132c5001L}
{Lacy}, M., {Baum}, S.~A., {Chandler}, C.~J., {et~al.} 2020, \pasp, 132, 035001, \dodoi{10.1088/1538-3873/ab63eb}

\bibitem[{{Lam} {et~al.}(2016){Lam}, {Cordes}, {Chatterjee}, {Jones}, {McLaughlin}, \& {Armstrong}}]{2016ApJ...821...66L}
{Lam}, M.~T., {Cordes}, J.~M., {Chatterjee}, S., {et~al.} 2016, \apj, 821, 66, \dodoi{10.3847/0004-637X/821/1/66}

\bibitem[{{Law} {et~al.}(2024){Law}, {Sharma}, {Ravi}, {Chen}, {Catha}, {Connor}, {Faber}, {Hallinan}, {Harnach}, {Hellbourg}, {Hobbs}, {Hodge}, {Hodges}, {Lamb}, {Rasmussen}, {Sherman}, {Shi}, {Simard}, {Squillace}, {Weinreb}, {Woody}, \& {Yurk}}]{2024ApJ...967...29L}
{Law}, C.~J., {Sharma}, K., {Ravi}, V., {et~al.} 2024, \apj, 967, 29, \dodoi{10.3847/1538-4357/ad3736}

\bibitem[{{Lorimer} {et~al.}(2007){Lorimer}, {Bailes}, {McLaughlin}, {Narkevic}, \& {Crawford}}]{2007Sci...318..777L}
{Lorimer}, D.~R., {Bailes}, M., {McLaughlin}, M.~A., {Narkevic}, D.~J., \& {Crawford}, F. 2007, Science, 318, 777, \dodoi{10.1126/science.1147532}

\bibitem[{{Lyutikov} {et~al.}(2016){Lyutikov}, {Burzawa}, \& {Popov}}]{2016MNRAS.462..941L}
{Lyutikov}, M., {Burzawa}, L., \& {Popov}, S.~B. 2016, \mnras, 462, 941, \dodoi{10.1093/mnras/stw1669}

\bibitem[{{Ma} {et~al.}(2019){Ma}, {Mao}, {Stil}, {Basu}, {West}, {Heiles}, {Hill}, \& {Betti}}]{2019MNRAS.487.3454M}
{Ma}, Y.~K., {Mao}, S.~A., {Stil}, J., {et~al.} 2019, \mnras, 487, 3454, \dodoi{10.1093/mnras/stz1328}

\bibitem[{{Macquart} {et~al.}(2012){Macquart}, {Ekers}, {Feain}, \& {Johnston-Hollitt}}]{2012ApJ...750..139M}
{Macquart}, J.~P., {Ekers}, R.~D., {Feain}, I., \& {Johnston-Hollitt}, M. 2012, \apj, 750, 139, \dodoi{10.1088/0004-637X/750/2/139}

\bibitem[{{Macquart} {et~al.}(2020){Macquart}, {Prochaska}, {McQuinn}, {Bannister}, {Bhandari}, {Day}, {Deller}, {Ekers}, {James}, {Marnoch}, {Os{\l}owski}, {Phillips}, {Ryder}, {Scott}, {Shannon}, \& {Tejos}}]{2020Natur.581..391M}
{Macquart}, J.~P., {Prochaska}, J.~X., {McQuinn}, M., {et~al.} 2020, \nat, 581, 391, \dodoi{10.1038/s41586-020-2300-2}

\bibitem[{{Manchester} {et~al.}(2005){Manchester}, {Hobbs}, {Teoh}, \& {Hobbs}}]{2005AJ....129.1993M}
{Manchester}, R.~N., {Hobbs}, G.~B., {Teoh}, A., \& {Hobbs}, M. 2005, \aj, 129, 1993, \dodoi{10.1086/428488}

\bibitem[{{Mannings} {et~al.}(2023){Mannings}, {Pakmor}, {Prochaska}, {van de Voort}, {Simha}, {Shannon}, {Tejos}, {Deller}, \& {Rafelski}}]{2023ApJ...954..179M}
{Mannings}, A.~G., {Pakmor}, R., {Prochaska}, J.~X., {et~al.} 2023, \apj, 954, 179, \dodoi{10.3847/1538-4357/ace7bb}

\bibitem[{{Mao} {et~al.}(2014){Mao}, {Banfield}, {Gaensler}, {Rudnick}, {Stil}, {Purcell}, {Beck}, {Farnes}, {O'Sullivan}, {Schnitzeler}, {Willis}, {Sun}, {Carretti}, {Dolag}, {Sokoloff}, {Kothes}, {Wolleben}, {Heald}, {Geisbuesch}, {Robishaw}, {Afonso}, {Magalh{\~a}es}, {Lundgren}, {Haverkorn}, {Oppermann}, \& {Taylor}}]{2014arXiv1401.1875M}
{Mao}, S.~A., {Banfield}, J., {Gaensler}, B., {et~al.} 2014, arXiv e-prints, arXiv:1401.1875, \dodoi{10.48550/arXiv.1401.1875}

\bibitem[{{Marcote} {et~al.}(2020){Marcote}, {Nimmo}, {Hessels}, {Tendulkar}, {Bassa}, {Paragi}, {Keimpema}, {Bhardwaj}, {Karuppusamy}, {Kaspi}, {Law}, {Michilli}, {Aggarwal}, {Andersen}, {Archibald}, {Bandura}, {Bower}, {Boyle}, {Brar}, {Burke-Spolaor}, {Butler}, {Cassanelli}, {Chawla}, {Demorest}, {Dobbs}, {Fonseca}, {Giri}, {Good}, {Gourdji}, {Josephy}, {Kirichenko}, {Kirsten}, {Landecker}, {Lang}, {Lazio}, {Li}, {Lin}, {Linford}, {Masui}, {Mena-Parra}, {Naidu}, {Ng}, {Patel}, {Pen}, {Pleunis}, {Rafiei-Ravandi}, {Rahman}, {Renard}, {Scholz}, {Siegel}, {Smith}, {Stairs}, {Vanderlinde}, \& {Zwaniga}}]{2020Natur.577..190M}
{Marcote}, B., {Nimmo}, K., {Hessels}, J.~W.~T., {et~al.} 2020, \nat, 577, 190, \dodoi{10.1038/s41586-019-1866-z}

\bibitem[{{Masui} {et~al.}(2015){Masui}, {Lin}, {Sievers}, {Anderson}, {Chang}, {Chen}, {Ganguly}, {Jarvis}, {Kuo}, {Li}, {Liao}, {McLaughlin}, {Pen}, {Peterson}, {Roman}, {Timbie}, {Voytek}, \& {Yadav}}]{2015Natur.528..523M}
{Masui}, K., {Lin}, H.-H., {Sievers}, J., {et~al.} 2015, \nat, 528, 523, \dodoi{10.1038/nature15769}

\bibitem[{{Mckinven} {et~al.}(2023){Mckinven}, {Gaensler}, {Michilli}, {Masui}, {Kaspi}, {Su}, {Bhardwaj}, {Cassanelli}, {Chawla}, {Dong}, {Fonseca}, {Leung}, {Li}, {Ng}, {Patel}, {Pearlman}, {Petroff}, {Pleunis}, {Rafiei-Ravandi}, {Rahman}, {Sand}, {Shin}, {Stairs}, \& {Tendulkar}}]{2023ApJ...951...82M}
{Mckinven}, R., {Gaensler}, B.~M., {Michilli}, D., {et~al.} 2023, \apj, 951, 82, \dodoi{10.3847/1538-4357/acd188}

\bibitem[{{Metzger} {et~al.}(2017){Metzger}, {Berger}, \& {Margalit}}]{2017ApJ...841...14M}
{Metzger}, B.~D., {Berger}, E., \& {Margalit}, B. 2017, \apj, 841, 14, \dodoi{10.3847/1538-4357/aa633d}

\bibitem[{{Metzger} {et~al.}(2019){Metzger}, {Margalit}, \& {Sironi}}]{2019MNRAS.485.4091M}
{Metzger}, B.~D., {Margalit}, B., \& {Sironi}, L. 2019, \mnras, 485, 4091, \dodoi{10.1093/mnras/stz700}

\bibitem[{{Mevius}(2018)}]{2018ascl.soft06024M}
{Mevius}, M. 2018, ASCL.
\newblock \url{https://ascl.net/1806.024}

\bibitem[{{Michilli} {et~al.}(2023){Michilli}, {Bhardwaj}, {Brar}, {Gaensler}, {Kaspi}, {Kirichenko}, {Masui}, {Mckinven}, {Ng}, {Patel}, {Sand}, {Scholz}, {Shin}, {Siegel}, {Stairs}, {Cassanelli}, {Cook}, {Dobbs}, {Dong}, {Fonseca}, {Ibik}, {Kaczmarek}, {Leung}, {Pearlman}, {Petroff}, {Pleunis}, {Rafiei-Ravandi}, {Sanghavi}, {Shaw}, \& {Tendulkar}}]{2023ApJ...950..134M}
{Michilli}, D., {Bhardwaj}, M., {Brar}, C., {et~al.} 2023, \apj, 950, 134, \dodoi{10.3847/1538-4357/accf89}

\bibitem[{{Mo} {et~al.}(2023){Mo}, {Zhu}, {Wang}, {Tang}, \& {Feng}}]{2023MNRAS.518..539M}
{Mo}, J.-F., {Zhu}, W., {Wang}, Y., {Tang}, L., \& {Feng}, L.-L. 2023, \mnras, 518, 539, \dodoi{10.1093/mnras/stac3104}

\bibitem[{{Ng} {et~al.}(2024){Ng}, {Pandhi}, {Mckinven}, {Curtin}, {Shin}, {Fonseca}, {Gaensler}, {Jow}, {Kaspi}, {Li}, {Main}, {Masui}, {Michilli}, {Nimmo}, {Pleunis}, {Scholz}, {Stairs}, {Bhardwaj}, {Brar}, {Cassanelli}, {Joseph}, {Pearlman}, {Rafiei-Ravandi}, \& {Smith}}]{2024arXiv241109045N}
{Ng}, C., {Pandhi}, A., {Mckinven}, R., {et~al.} 2024, arXiv e-prints, arXiv:2411.09045, \dodoi{10.48550/arXiv.2411.09045}

\bibitem[{{Nimmo} {et~al.}(2022){Nimmo}, {Hessels}, {Kirsten}, {Keimpema}, {Cordes}, {Snelders}, {Hewitt}, {Karuppusamy}, {Archibald}, {Bezrukovs}, {Bhardwaj}, {Blaauw}, {Buttaccio}, {Cassanelli}, {Conway}, {Corongiu}, {Feiler}, {Fonseca}, {Forss{\'e}n}, {Gawro{\'n}ski}, {Giroletti}, {Kharinov}, {Leung}, {Lindqvist}, {Maccaferri}, {Marcote}, {Masui}, {Mckinven}, {Melnikov}, {Michilli}, {Mikhailov}, {Ng}, {Orbidans}, {Ould-Boukattine}, {Paragi}, {Pearlman}, {Petroff}, {Rahman}, {Scholz}, {Shin}, {Smith}, {Stairs}, {Surcis}, {Tendulkar}, {Vlemmings}, {Wang}, {Yang}, \& {Yuan}}]{2022NatAs...6..393N}
{Nimmo}, K., {Hessels}, J.~W.~T., {Kirsten}, F., {et~al.} 2022, Nature Astronomy, 6, 393, \dodoi{10.1038/s41550-021-01569-9}

\bibitem[{{Niu} {et~al.}(2022){Niu}, {Aggarwal}, {Li}, {Zhang}, {Chatterjee}, {Tsai}, {Yu}, {Law}, {Burke-Spolaor}, {Cordes}, {Zhang}, {Ocker}, {Yao}, {Wan}, {Feng}, {Niino}, {Bochenek}, {Cruces}, {Connor}, {Jiang}, {Dai}, {Luo}, {Li}, {Miao}, {Niu}, {Anna-Thomas}, {Sydnor}, {Stern}, {Wang}, {Yuan}, {Yue}, {Zhou}, {Yan}, {Zhu}, \& {Zhang}}]{2022Natur.606..873N}
{Niu}, C.~H., {Aggarwal}, K., {Li}, D., {et~al.} 2022, \nat, 606, 873, \dodoi{10.1038/s41586-022-04755-5}

\bibitem[{{Oppermann} {et~al.}(2012){Oppermann}, {Junklewitz}, {Robbers}, {Bell}, {En{\ss}lin}, {Bonafede}, {Braun}, {Brown}, {Clarke}, {Feain}, {Gaensler}, {Hammond}, {Harvey-Smith}, {Heald}, {Johnston-Hollitt}, {Klein}, {Kronberg}, {Mao}, {McClure-Griffiths}, {O'Sullivan}, {Pratley}, {Robishaw}, {Roy}, {Schnitzeler}, {Sotomayor-Beltran}, {Stevens}, {Stil}, {Sunstrum}, {Tanna}, {Taylor}, \& {Van Eck}}]{2012A&A...542A..93O}
{Oppermann}, N., {Junklewitz}, H., {Robbers}, G., {et~al.} 2012, \aap, 542, A93, \dodoi{10.1051/0004-6361/201118526}

\bibitem[{{Oppermann} {et~al.}(2015){Oppermann}, {Junklewitz}, {Greiner}, {En{\ss}lin}, {Akahori}, {Carretti}, {Gaensler}, {Goobar}, {Harvey-Smith}, {Johnston-Hollitt}, {Pratley}, {Schnitzeler}, {Stil}, \& {Vacca}}]{2015A&A...575A.118O}
{Oppermann}, N., {Junklewitz}, H., {Greiner}, M., {et~al.} 2015, \aap, 575, A118, \dodoi{10.1051/0004-6361/201423995}

\bibitem[{{Osinga} {et~al.}(2024){Osinga}, {van Weeren}, {Rudnick}, {Andrade-Santos}, {Bonafede}, {Clarke}, {Duncan}, {Giacintucci}, \& {R{\"o}ttgering}}]{2024arXiv240807178O}
{Osinga}, E., {van Weeren}, R.~J., {Rudnick}, L., {et~al.} 2024, arXiv e-prints, arXiv:2408.07178, \dodoi{10.48550/arXiv.2408.07178}

\bibitem[{{O'Sullivan} {et~al.}(2023){O'Sullivan}, {Shimwell}, {Hardcastle}, {Tasse}, {Heald}, {Carretti}, {Br{\"u}ggen}, {Vacca}, {Sobey}, {Van Eck}, {Horellou}, {Beck}, {Bilicki}, {Bourke}, {Botteon}, {Croston}, {Drabent}, {Duncan}, {Heesen}, {Ideguchi}, {Kirwan}, {Lawlor}, {Mingo}, {Nikiel-Wroczy{\'n}ski}, {Piotrowska}, {Scaife}, \& {van Weeren}}]{2023MNRAS.519.5723O}
{O'Sullivan}, S.~P., {Shimwell}, T.~W., {Hardcastle}, M.~J., {et~al.} 2023, \mnras, 519, 5723, \dodoi{10.1093/mnras/stac3820}

\bibitem[{{Pandhi} {et~al.}(2022){Pandhi}, {Hutschenreuter}, {West}, {Gaensler}, \& {Stock}}]{2022MNRAS.516.4739P}
{Pandhi}, A., {Hutschenreuter}, S., {West}, J.~L., {Gaensler}, B.~M., \& {Stock}, A. 2022, \mnras, 516, 4739, \dodoi{10.1093/mnras/stac2314}

\bibitem[{{Pandhi} {et~al.}(2024){Pandhi}, {Pleunis}, {Mckinven}, {Gaensler}, {Su}, {Ng}, {Bhardwaj}, {Brar}, {Cassanelli}, {Cook}, {Curtin}, {Kaspi}, {Lazda}, {Leung}, {Li}, {Masui}, {Michilli}, {Nimmo}, {Pearlman}, {Petroff}, {Rafiei-Ravandi}, {Sand}, {Scholz}, {Shin}, {Smith}, \& {Stairs}}]{2024ApJ...968...50P}
{Pandhi}, A., {Pleunis}, Z., {Mckinven}, R., {et~al.} 2024, \apj, 968, 50, \dodoi{10.3847/1538-4357/ad40aa}

\bibitem[{{Piro} \& {Gaensler}(2018)}]{2018ApJ...861..150P}
{Piro}, A.~L., \& {Gaensler}, B.~M. 2018, \apj, 861, 150, \dodoi{10.3847/1538-4357/aac9bc}

\bibitem[{{Price} {et~al.}(2021){Price}, {Flynn}, \& {Deller}}]{2021PASA...38...38P}
{Price}, D.~C., {Flynn}, C., \& {Deller}, A. 2021, \pasa, 38, e038, \dodoi{10.1017/pasa.2021.33}

\bibitem[{{Purcell} {et~al.}(2020){Purcell}, {Van Eck}, {West}, {Sun}, \& {Gaensler}}]{2020ascl.soft05003P}
{Purcell}, C.~R., {Van Eck}, C.~L., {West}, J., {Sun}, X.~H., \& {Gaensler}, B.~M. 2020, ASCL, 2005.003.
\newblock \doeprint{2005.003}

\bibitem[{{Ravi} {et~al.}(2016){Ravi}, {Shannon}, {Bailes}, {Bannister}, {Bhandari}, {Bhat}, {Burke-Spolaor}, {Caleb}, {Flynn}, {Jameson}, {Johnston}, {Keane}, {Kerr}, {Tiburzi}, {Tuntsov}, \& {Vedantham}}]{2016Sci...354.1249R}
{Ravi}, V., {Shannon}, R.~M., {Bailes}, M., {et~al.} 2016, Science, 354, 1249, \dodoi{10.1126/science.aaf6807}

\bibitem[{{Schnitzeler}(2010)}]{2010MNRAS.409L..99S}
{Schnitzeler}, D.~H.~F.~M. 2010, \mnras, 409, L99, \dodoi{10.1111/j.1745-3933.2010.00957.x}

\bibitem[{{Shannon} {et~al.}(2024){Shannon}, {Bannister}, {Bera}, {Bhandari}, {Day}, {Deller}, {Dial}, {Dobie}, {Ekers}, {Fong}, {Glowacki}, {Gordon}, {Gourdji}, {Jaini}, {James}, {Kumar}, {Mahony}, {Marnoch}, {Muller}, {Prochaska}, {Qiu}, {Ryder}, {Sadler}, {Scott}, {Tejos}, {Uttarkar}, \& {Wang}}]{2024arXiv240802083S}
{Shannon}, R.~M., {Bannister}, K.~W., {Bera}, A., {et~al.} 2024, arXiv e-prints, arXiv:2408.02083, \dodoi{10.48550/arXiv.2408.02083}

\bibitem[{{Sharma} {et~al.}(2024){Sharma}, {Ravi}, {Connor}, {Law}, {Ocker}, {Sherman}, {Kosogorov}, {Faber}, {Hallinan}, {Harnach}, {Hellbourg}, {Hobbs}, {Hodge}, {Hodges}, {Lamb}, {Rasmussen}, {Somalwar}, {Weinreb}, {Woody}, {Leja}, {Anand}, {Das}, {Qin}, {Rose}, {Dong}, {Miller}, \& {Yao}}]{2024Natur.635...61S}
{Sharma}, K., {Ravi}, V., {Connor}, L., {et~al.} 2024, \nat, 635, 61, \dodoi{10.1038/s41586-024-08074-9}

\bibitem[{{Sherman} {et~al.}(2024){Sherman}, {Connor}, {Ravi}, {Law}, {Chen}, {Catha}, {Faber}, {Hallinan}, {Harnach}, {Hellbourg}, {Hobbs}, {Hodge}, {Hodges}, {Lamb}, {Rasmussen}, {Sharma}, {Shi}, {Simard}, {Somalwar}, {Squillace}, {Weinreb}, {Woody}, {Yadlapalli}, \& {The Deep Synoptic Array team}}]{2024ApJ...964..131S}
{Sherman}, M.~B., {Connor}, L., {Ravi}, V., {et~al.} 2024, \apj, 964, 131, \dodoi{10.3847/1538-4357/ad275e}

\bibitem[{{Sobey} {et~al.}(2019){Sobey}, {Bilous}, {Grie{\ss}meier}, {Hessels}, {Karastergiou}, {Keane}, {Kondratiev}, {Kramer}, {Michilli}, {Noutsos}, {Pilia}, {Polzin}, {Stappers}, {Tan}, {van Leeuwen}, {Verbiest}, {Weltevrede}, {Heald}, {Alves}, {Carretti}, {En{\ss}lin}, {Haverkorn}, {Iacobelli}, {Reich}, \& {Van Eck}}]{2019MNRAS.484.3646S}
{Sobey}, C., {Bilous}, A.~V., {Grie{\ss}meier}, J.~M., {et~al.} 2019, \mnras, 484, 3646, \dodoi{10.1093/mnras/stz214}

\bibitem[{{Taylor} {et~al.}(2009){Taylor}, {Stil}, \& {Sunstrum}}]{2009ApJ...702.1230T}
{Taylor}, A.~R., {Stil}, J.~M., \& {Sunstrum}, C. 2009, \apj, 702, 1230, \dodoi{10.1088/0004-637X/702/2/1230}

\bibitem[{{Thomson} {et~al.}(2023){Thomson}, {McConnell}, {Lenc}, {Galvin}, {Rudnick}, {Heald}, {Hale}, {Duchesne}, {Anderson}, {Carretti}, {Federrath}, {Gaensler}, {Harvey-Smith}, {Haverkorn}, {Hotan}, {Ma}, {Murphy}, {McClure-Griffiths}, {Moss}, {O'Sullivan}, {Raja}, {Seta}, {Van Eck}, {West}, {Whiting}, \& {Wieringa}}]{2023PASA...40...40T}
{Thomson}, A. J.~M., {McConnell}, D., {Lenc}, E., {et~al.} 2023, \pasa, 40, e040, \dodoi{10.1017/pasa.2023.38}

\bibitem[{{Van Eck} {et~al.}(2023){Van Eck}, {Gaensler}, {Hutschenreuter}, {Livingston}, {Ma}, {Riseley}, {Thomson}, {Adebahr}, {Basu}, {Birkinshaw}, {En{\ss}lin}, {Heald}, {Mao}, \& {McClure-Griffiths}}]{2023ApJS..267...28V}
{Van Eck}, C.~L., {Gaensler}, B.~M., {Hutschenreuter}, S., {et~al.} 2023, \apjs, 267, 28, \dodoi{10.3847/1538-4365/acda24}

\bibitem[{{Vanderwoude} {et~al.}(2024){Vanderwoude}, {West}, {Gaensler}, {Rudnick}, {Van Eck}, {Thomson}, {Andernach}, {Anderson}, {Carretti}, {Heald}, {Leahy}, {McClure-Griffiths}, {O'Sullivan}, {Tahani}, \& {Willis}}]{2024AJ....167..226V}
{Vanderwoude}, S., {West}, J.~L., {Gaensler}, B.~M., {et~al.} 2024, \aj, 167, 226, \dodoi{10.3847/1538-3881/ad2fc8}

\bibitem[{Virtanen {et~al.}(2020)Virtanen, Gommers, Oliphant, Haberland, Reddy, Cournapeau, Burovski, Peterson, Weckesser, Bright, {van der Walt}, Brett, Wilson, Millman, Mayorov, Nelson, Jones, Kern, Larson, Carey, Polat, Feng, Moore, {VanderPlas}, Laxalde, Perktold, Cimrman, Henriksen, Quintero, Harris, Archibald, Ribeiro, Pedregosa, {van Mulbregt}, \& {SciPy 1.0 Contributors}}]{2020SciPy-NMeth}
Virtanen, P., Gommers, R., Oliphant, T.~E., {et~al.} 2020, Nature Methods, 17, 261, \dodoi{10.1038/s41592-019-0686-2}

\bibitem[{{Yamasaki} \& {Totani}(2020)}]{2020ApJ...888..105Y}
{Yamasaki}, S., \& {Totani}, T. 2020, \apj, 888, 105, \dodoi{10.3847/1538-4357/ab58c4}

\end{thebibliography}
\bibliographystyle{aasjournal}

\end{document}